\documentclass{aa}  
\usepackage{graphicx}
\usepackage{txfonts}
\begin{document}

   \title{Transient effects in Herschel/PACS spectroscopy}
   \author{Dario Fadda
          \inst{1}
          \fnmsep\thanks{Moving to the SOFIA Science Center (USRA)}
          \and
          Jeffery D. Jacobson\inst{2}
          \and
          Philip N. Appleton\inst{2}
          }

   \institute{Instituto de Astrofisica de Canarias, E-38205, La Laguna, Tenerife, Spain\\
             Universidad de La Laguna, Dpto. de Astrofísica, E-38206 La Laguna, Tenerife, Spain\\
         \and
         NASA Herschel Science Center --- California Institute of Technology,
         MC 100-22, Pasadena, CA 91125, USA
         \\
             \email{jdj@ipac.caltech.edu,apple@ipac.caltech.edu}
}

 
  \abstract
      { The Ge:Ga detectors used in the PACS spectrograph onboard the
        {\it Herschel} space telescope react to changes of the
        incident flux with a certain delay. This generates transient
        effects on the resulting signal which can be important and
        last for up to an hour.}
      { The paper presents a study of the effects of transients on the detected signal
        and proposes methods to mitigate them especially in the case of the ``unchopped'' mode.}
      { Since transients can arise from a
      variety of causes, we classified them in three main
      categories: transients caused by sudden variations of the
      continuum due to the observational mode used; transients caused
      by cosmic ray impacts on the detectors; transients caused by a
      continuous smooth variation of the continuum during a wavelength
      scan.  We propose a method to disentangle these effects and
      treat them separately.  In particular, we show that a linear
      combination of three exponential functions is needed to fit the
      response variation of the detectors during a transient.  An
      algorithm to detect, fit, and correct transient effects is
      presented.  }
      { The solution proposed to correct the
      signal for the effects of transients substantially improves the
      quality of the final reduction with respect to the standard methods
      used for archival reduction in the cases where transient effects are most pronounced.  }
      { The
      programs developed to implement the corrections are offered
      through two new interactive data reduction pipelines in the
      latest releases of the Herschel Interactive Processing
      Environment.  }

      \keywords{
        methods: data analysis --- techniques: spectroscopic --- infrared:
        general}

   \maketitle
%

\section{Introduction}

The {\it Herschel} space observatory \citep{2010A&A...518L...1P}
completed its mission on April 29, 2013 after performing a total of
23,400 hours of scientific observations during almost 4 years of
activity ( 1446 operational days). One of the most used instruments
aboard {\it Herschel} was the PACS spectrometer
\citep{2010A&A...518L...2P}.  Approximately one quarter of the total
scientific time was devoted to PACS spectroscopy. Although most
of the observations were performed in the standard ``chop-nod''
mode, a substantial fraction (30\% of the observations, corresponding
to the 25\% of the total spectroscopy time) used two alternative
modes.  The ``wavelength switching'' mode was released to users after
the start of the mission to allow PACS spectrometer observations to be
made in crowded fields where chopping was not possible. A year later,
this mode was replaced by the so-called ``unchopped'' mode. By the end
of the {\it Herschel} operation mission, this mode was used for
approximately one-third of all PACS spectroscopic observations.  The
primary focus of this paper is to describe an optimal way to reduce
data taken in the ``unchopped'' mode.

The PACS spectrometer was able to observe spectroscopically between
60~$\mu$m and 210~$\mu$m using Ge:Ga photo-conductor arrays. This type of
detectors suffers from systematic memory effects of the response which
can bias the photometry of sources and increase the noise in the
signal. Such effects have been documented and studied for similar
detectors on previous space observatories such as ISOCAM \citep[][]{2000A&AS..141..533C,2001MNRAS.325.1173L} and
MIPS \citep[][Fig. 17]{2006AJ....131.2859F}.

In this paper we briefly review the observational modes of the PACS
spectrograph and the data reduction techniques. We describe how the
``unchopped'' mode is particularly sensitive to sudden changes in
incident flux, and cosmic ray glitches, which cause memory effects in
the detector responses.  Finally, we introduce the technique used to
correct most of these effects, and show a few selected examples. The
methods and software described in the paper are now implemented in the
Herschel Interactive Processing Environment
(HIPE\footnote{www.cosmos.esa.int/web/herschel/hipe-download})
\citep[][]{2010ASPC..434..139O}.  The comparisons presented in the
paper are between the archived SPG (standard product generation)
products, generated with HIPE 14.0 and calibration data version 72,
and our pipeline in HIPE 15.

\section{The PACS spectrometer}

We recommend reading \citet{2010A&A...518L...2P} and the PACS
observational manual for a detailed description of the PACS
spectrometer. As a way to introduce some terminology used in the
paper, we give here a concise description of the instrument and how it
works.

The PACS spectrometer was an IFU (integral field unit) composed by a
matrix of $5\times5$ spatial pixels (spaxels or space modules)
covering a field of $47'\times47'$ square arcminutes.  The $5\times5$
pixel image passed into an image slicer, which rearranged the $5\times5$
two-dimensional image into one-dimension ($1\times25$), which then fed a
Littrow-mounted diffraction grating where it operated at $1^{st}$, $2^{nd}$, and
$3^{rd}$ order.  The first order (red) and second or third order (blue)
were separated with a dichroic beam-splitter, where the spectra were
re-imaged onto separate detectors. The wavelength range covered is
51-105~$\mu$m for the blue (choice of 51-73 and 71-105~$\mu$m for the
$3^{rd}$ and $2^{nd}$ order) and 102-220~$\mu$m for the red, respectively.

The dispersed light was detected by two (low and high-stressed) Ge:Ga
photo-conductors arrays with $25\times16$ pixels.  In the following we
call spectral pixels the individual pixels of the two
arrays. We call ``module'' or ``spaxel'' (from spatial pixel) the set
of 16 spectral pixels corresponding to the dispersed light from a
single patch of $9.4\times9.4$ square arcseconds on the sky.

Since the instantaneous wavelength range covered by 16 spectral pixels
is small ($\sim$1500~km/s for many observations), the grating was
typically stepped through a range of grating angles during an
observation. This operation is referred as a ``wavelength scan''.
During a standard observation, the desired wavelength range is covered
by moving the grating back and forth.  These movements are called up-
and down-scans, since the wavelength seen by a single spectral pixel
first increases and then decreases as the grating executes a scan
first in one direction, and then back.

The other mobile part of the system is the chopper which lies at the
entrance to the whole PACS instrument. This is a mirror which allows
the IFU to point at different parts of the sky.  In standard chop-nod
mode, the chopping mirror is used to alternatively point rapidly to
the source and then a background position, while maintaining the same
telescope pointing.  The largest chopper throw available is
6~arcminutes, which is a limitation of the chop-nod mode.  If the
chopper mirror is moved to larger angles, it can access two internal
calibrators (BB1 and BB2) which are used for calibration during the
slew to the target source.

Since the mirror is passively cooled, the emission of the telescope is
significant, and typically dominates the total signal. For instance,
in the red, the emission seen by a given pixel at 120~$\mu$m is around
300~Jy.  This high telescope background level, although not generally
desirable in IR astronomy, has the advantage that it helps to mitigate
transient effects in the detectors by helping to keep the signal level
constant at the detectors.  A second property of the dominant
telescope signal is that it is always present, and can be used as a
relatively constant reference flux.  This property can be positively
exploited in the analysis of these PACS data.

\section{Observational modes}

\begin{figure}
  \hbox{
    \includegraphics[width=0.45\columnwidth]{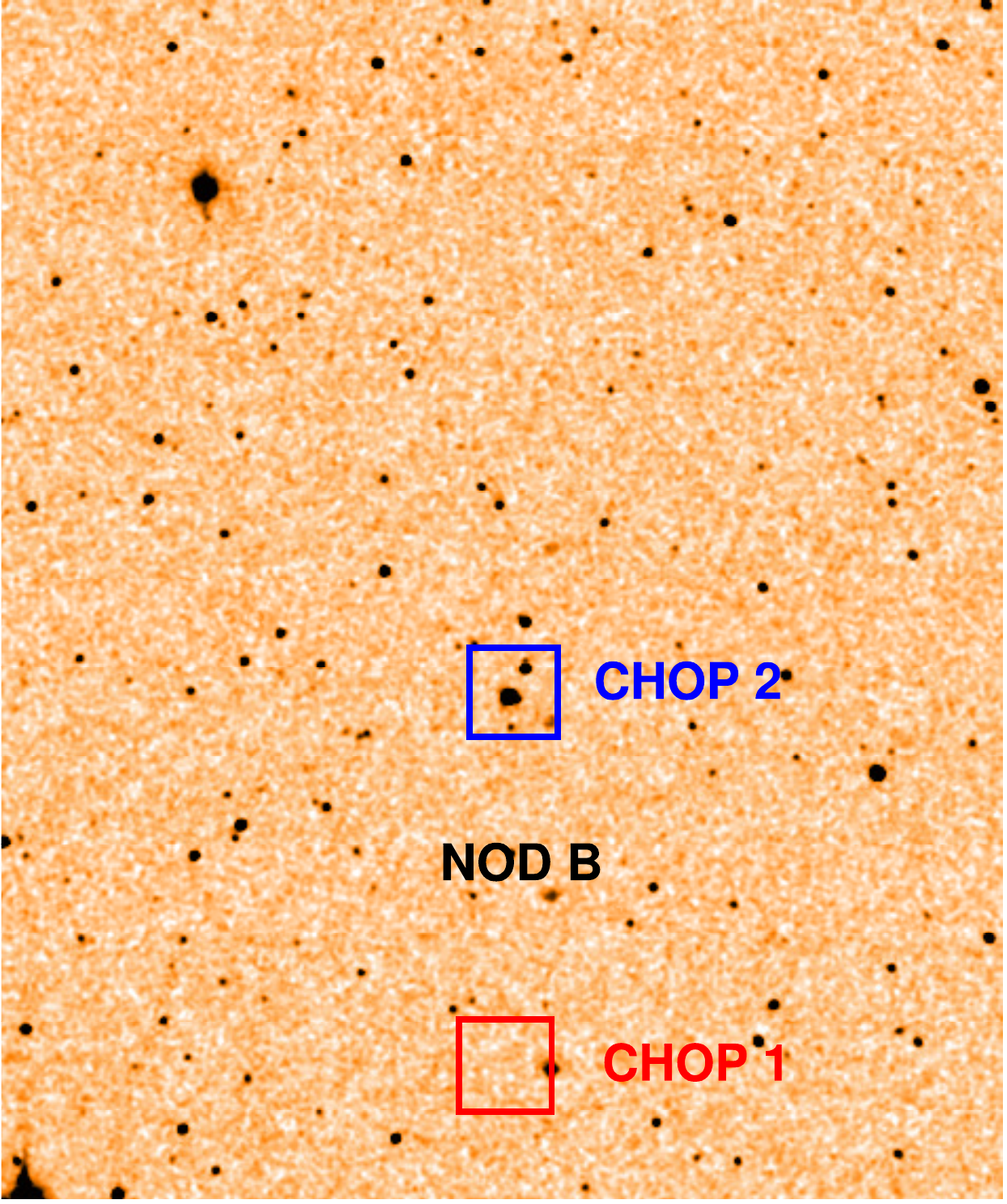}
    \includegraphics[width=0.45\columnwidth]{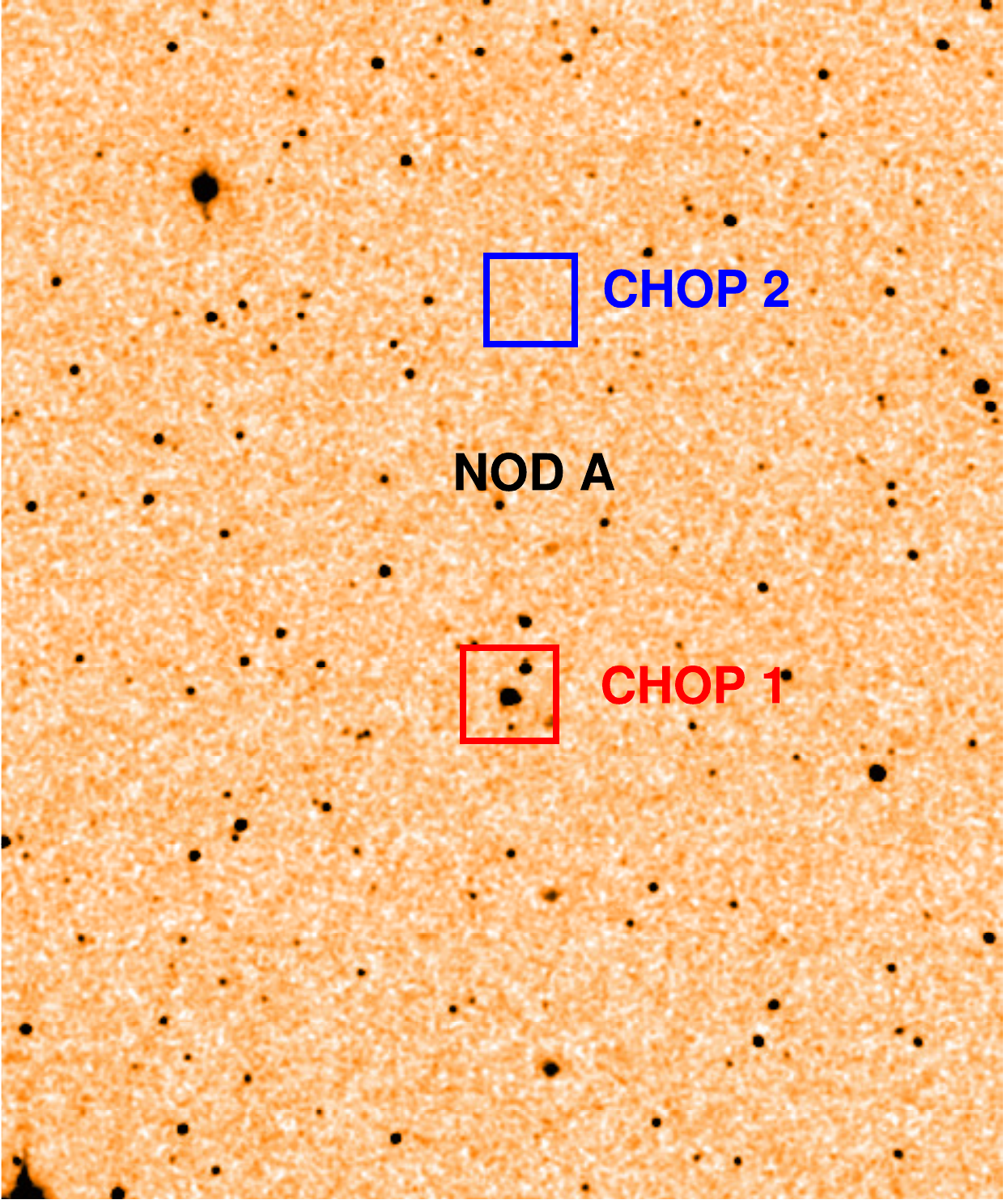}
  }
  \includegraphics[width=0.9\columnwidth]{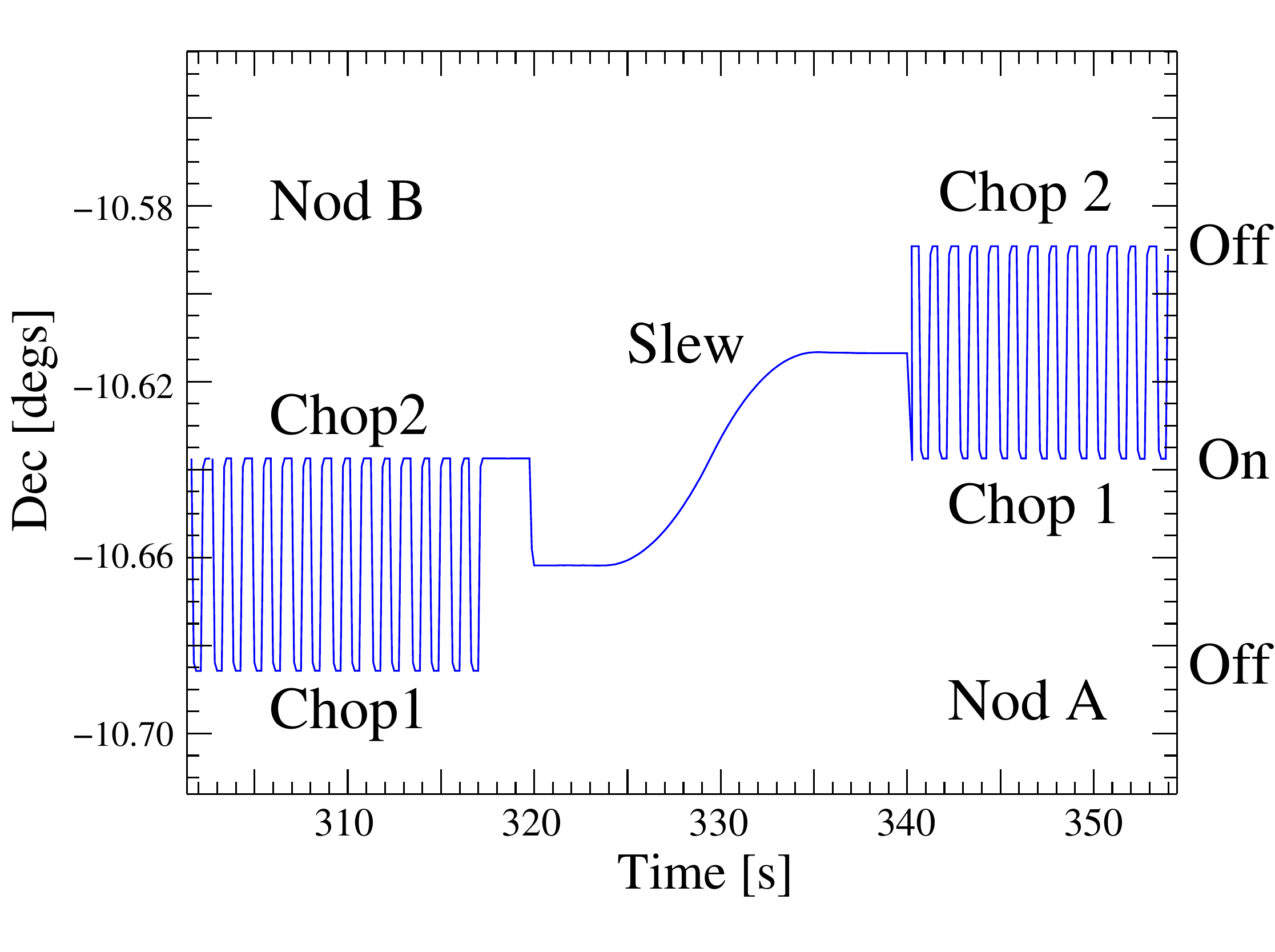}
  \caption{Example of chop-nod observation footprints (top), and
    observed position projected on the sky (bottom). In the first nod,
    the detector alternates between two optical paths to observe the
    object and the background. These paths have slightly different
    telescope background signals. In the second nod, the optical paths
    are inverted after slewing the telescope. By adding the two
    object-background differences, it is possible to remove the
    effects of the different telescope background levels.  }
  \label{fig:chopnod}
\end{figure}

   Spectroscopic observations consist of a calibration block
  followed by a series of science blocks. The calibration block is
  performed when the telescope is slewing to reach the target and
  involves rapid chopping between the two internal reference
  black-bodies.  The science target is then observed in one or more
  bands using either the ``chop-nod'' or the ``unchopped''
  observational mode.  The ``chop-nod'' mode is used for isolated
  sources and involves the continuous chopping between target and an
  ``off'' position.  The ``unchopped mode'' is used to observe crowded
  fields or extended sources.  It consists of a staring observation of
  the science target followed by an observation of the reference
  ``off'' position by moving the telescope. This mode replaced the
  ``wavelength-switching'' mode used for earlier observations.
  Instead of chopping, the ``wavelength switching'' used a wavelength
  modulation to move the line on the detector array by a wavelength
  equivalent to the FWHM of the line. This allowed the differential
  spectrum to be measured, but it was found to be inefficient during
  the verification phase and deprecated.  Observations in ``unchopped
mode'' suffered from detector transient effects. This paper is
directed at ways of solving some of these problems within the HIPE
software.

\subsection{Chop-nod mode}

The standard way to observe with the PACS spectrometer  was the
``chop-nod'' mode. During a wavelength scan, the chopper modulates
between an ``on-source'' and ``off-source'' position. In this way, any
variation of the response slower than the chopping time can be
subtracted using the ``off-source'' constant reference. However, since
the on- and off-source chop signals follow different optical paths
through the telescope, the observed telescope background emission is
slightly different in the two chop positions.  To remove this effect,
the telescope is nodded. In the left panel of Figure~\ref{fig:chopnod}
we show the first nod position (nod B) when the source is observed in
the chop 2 position. The right-hand panel shows the situation after a
small slew which places the source in chop position 1 where the
observation is repeated (nod A).  The optical paths used to observe
source and off-field are inverted so that the average of the ``source
minus off'' signals is largely unaffected by the different telescope
backgrounds.

Two types of submodes are defined in the chop-nod mode. In the case
the observation involves an unresolved line, the ``line chop-nod''
mode defines an optimal way to observe the line and the surrounding
continuum.  If a larger wavelength range is required, the ``range
chop-nod'' mode can be used to manually enter the range of wavelengths
to be scanned by the grating.

The data reduction is done on each individual pixel.  The simplest
reduction technique uses the calibration block to compute the average
response during the observation. The response is defined for each pixel
as the ratio between the measured and expected signal from the internal
calibrators which was measured in the laboratory.
The measured signal is composed of the telescope
background ($T_A$ and $T_B$ in the two chopping positions), source
flux ($s$), and dark ($d$) signal.  The telescope background is
dominated by the emission of the primary mirror, and depends on the
temperature of the spot which is seen by the detector. Because of
differential variations in the mirror temperature, and its emissivity
with position on the mirror, the different chopper positions tend to
see differing degrees of telescope background. Although, during a
given observation, the overall temperature of the mirror changed
slowly, it was measured to vary due to changes in illumination of the
spacecraft by the Sun on longer timescales than a single observation.

For the measured signal $C$, in $V/s$, in a chop-nod observation we have: 
\begin{equation}
  \begin{array}{c@{\:=\:}r@{\quad}c@{\:=\:}r}
    C_1 & (T_A + d + s) \cdot R(t) &  C_2  &  (T_B + d) \cdot R(t) \\[2ex]
    C'_1 & (T_A + d)\cdot  R(t) & C'_2  &  (T_B + d + s)\cdot  R(t)
  \end{array}
\end{equation}
for the two chop positions in the first ($C_1$ and $C_2$) and second
nods ($C'_1$ and $C'_2$), respectively.  Here $R(t) = \rho_{\lambda}
\cdot r(t)$ is the product of the relative spectral response
function (RSRF) $\rho_{\lambda}$ and the response function $r(t)$ (V
s$^{-1}$ Jy$^{-1}$) which can vary with time.  T$_A$ and T$_B$
correspond to the signal detected from the telescope background at
the chopper positions A and B, respectively.

The simplest technique consists in estimating the signal source by
computing the differential signal, $\hat{s}$, between the chopping
positions and combining the nods:
\begin {equation}
  \hat{s} = [(C_1-C_2)+(C'_1-C'_2)]/(2\cdot \rho_{\lambda}\cdot r_{CB}), ~~~~~~~[Jy]
\end{equation}
where $r_{CB} = \left<r(t)\right>$ is the average response estimated
from the calibration block.  As we can see, the method works if $r(t)$
is approximately constant during a chopping period (1 sec) so that the
subtraction cancels the effect of a transient.  This approach was used
by the so called SPG (standard product generation) pipeline to
populate the Herschel Science Archive (HSA) for chop-nod
observations\footnote{http://www.cosmos.esa.int/web/herschel/science-archive}
until recently.

Since SPG 13, an alternative approach exploiting the knowledge of the telescope
background is used. This technique does not require the application of the
RSRF and response corrections since it uses the ratios between the
observations in the two chopping directions.  Nevertheless, the
accuracy of the results depends on the knowledge of the mean telescope
background. Also, ratios introduce more noise in the final signal with
respect to the standard reduction.

In formulae, if we define normalization as:
\begin{equation}
    N = \frac{C_1-C_2}{C_1+C_2} + \frac{C'_1-C'_2}{C'_1+C'_2}=
    \frac{2s}{T_A+T_B+2d+s},
\end{equation}
the signal normalized to the average telescope background can be expressed as:
\begin{equation} 
  \frac{s}{\left<T\right>+d} = \frac{s}{(T_A+T_B)/2+d} = \frac{N}{1-N/2}.
\end{equation}
The source flux can be therefore derived by multiplying the normalized
signal by the telescope background (dark included). This method frees
the result from the effects of the variable response although it
requires the knowledge of the dark. Since PACS has no shutter, the
dark was measured in the laboratory before the launch of {\it
  Herschel}. It is not known with great accuracy although its value is
negligible with respect to the telescope background.  The telescope
background was derived in flight by calibrating the emission of the
mirror with observed spectra of bright asteroids.

\subsection{Unchopped mode}

At the end of the verification phase, the ``wavelength switching''
mode \citep[see][]{2010A&A...518L...2P} was
released\footnote{herschel.esac.esa.int/Docs/AOTsReleaseStatus/\\ PACS\_WaveSwitching\_ReleaseNote\_20Jan2010.pdf}
and used in a few key programs. Several disadvantages became apparent
with this mode during its early use, leading to its replacement with
the ``unchopped'' mode. For example, by construction, the continuum
of the source could not be measured. Moreover, only observations of
unresolved lines could be measured correctly. Asymmetric lines or
lines broader than the wavelength switching interval were found to be
difficult to recover. For large wavelength ranges, SED shape
variations can lead to peculiar baselines in the final differential
spectrum. This made observations over large wavelength ranges
challenging.  Finally, the rapid switching of the grating between two
distant wavelength positions created mechanical oscillations which
required a long time to damp. This effect was more severe in space
than during ground testing, leading to longer time intervals between
useful observational samples and thus preventing an efficient removal of
rapid response variations.

It appeared clear that performing a slow grating scan while staring at
an object, followed by a similar observation pointed to an ``off''
position, yielded superior results compared with
``wavelength-switching''.  This ``unchopped'' mode had the advantage
of allowing large wavelength range scans, such as far-IR SEDs of
confused or extended regions.  Unlike the ``chop-nod'' mode, where the
two chopper positions sample different optical paths and therefore
different mirror temperatures, in the ``unchopped'' mode the optical
paths used for on- and off-source observations are the
same. Therefore, the mirror temperature in the on- and off-source are
exactly the same.  Moreover, in the case of chop-nod, the footprint of
the detector on the sky rotates slightly between the two chop
positions (see the PACS observer manual, Figure 4.7).  The effect,
though small, is worse for the larger chopper throws.  This means that
in the two nod positions, the only spaxel seeing precisely the same
part of the source is the central one. Spaxels further from the center
become successively mismatched in the two nods.  So, the reduction
works best for the central spaxel.  This problem does not exist for
the ``unchopped'' mode because the observation of the on- and
off-positions are made with a fixed central chopper position.

\begin{figure*}
  \begin{center}
    \includegraphics[width=1.7\columnwidth]{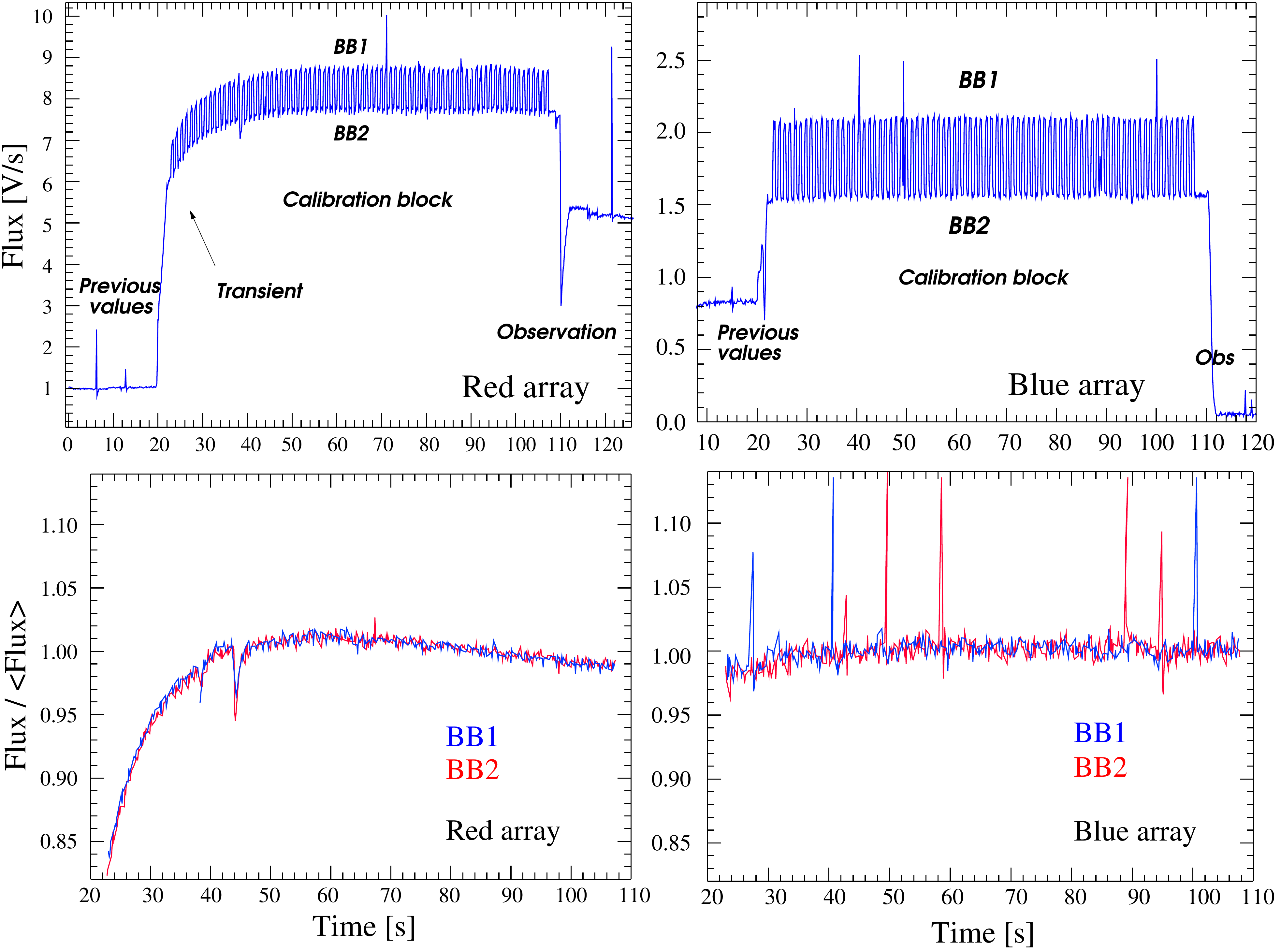}
  \end{center}
\caption{Transients affecting the calibration block for the red (left)
  and blue (right) arrays. During the calibration block, the detectors
  see alternatively the two blackbodies (top panels). Since the
  initial flux is significantly different from the fluxes of the
  internal blackbodies, a strong transient occurs during the first
  minute of the observation.  The signals normalized to the median
  values of each blackbody (bottom panels) show how the same transient
  affects the fluxes from the two blackbodies (blue and red
  lines). The transient effect is more pronounced for the red
  array. \label{fig:calblock} }
\end{figure*}

The ``unchopped'' line and range modes were released on September
2010, superseding the previous ``wavelength switching''
mode\footnote{herschel.esac.esa.int/Docs/AOTsReleaseStatus/\\ PACS\_Unchopped\_ReleaseNote\_20Sep2010.pdf}.
A specific ``unchopped'' mode for bright lines was released on April
2011\footnote{herschel.esac.esa.int/twiki/pub/Public/PacsAotReleaseNotes/\\ PACS\_UnchoppedReleaseNote\_BrightLines\_15Apr2011.pdf\\
Note that all the AOT release notes will be provided on the Herschel Legacy Library website starting from 2017.}.

In ``unchopped'' mode, a complete wavelength scan is done first on
the ``on-source'' (hereafter ON) position and then later on a
``off-source'' (hereafter OFF) position clear of source emission.
Unfortunately, variations of the response during the ON scan
cannot be corrected using the OFF scan because the two scans
are performed at different times. So, although unchopped
observations offer some advantages over the chop-nod mode, the effects
of transients require mitigation.  In this paper, we show the
typical transients found in the signal, and some techniques to model
and subtract them.

Three different sub-modes exist for the unchopped mode:
\begin{itemize}
\item {\it unchopped line},  for single line observations.
\item {\it unchopped bright line},  for single bright lines.
  It is 30\% more time efficient than the standard line mode since
  bright lines require less continuum to compute the line intensity.
\item {\it unchopped range}, for observations of the continuum
  or a complex of lines.
\end{itemize}
A further difference between these sub-modes is that the reference OFF
is observed during the same AOR (Astronomical Observation Request) in
the case of the ``unchopped line'', but has to be provided as a
separate AOR in the ``unchopped range''. In order to ensure the
observations are performed sequentially, the OFF observation in the
unchopped range scan is concatenated to the main observation within the
observation sequence.

\section{Transients}

 The term ``transient'' refers to a delayed response of a detector to
the variation of the incident flux. Transients are particularly
evident after sudden variations of flux on detector arrays \citep[see,
  e.g., ][]{2000A&AS..141..533C}.  In Figure~\ref{fig:calblock} we
show the signal detected during the calibration block when the chopper
points alternatively between the two internal black bodies (which have
different temperatures).  Passing from the previous observation to the
internal black bodies causes a clear transient effect which is more
important in the case of the red array (left panels). When the signal
is normalized to the asymptotic flux (bottom panels), the response
variation becomes evident.  It is interesting to note (left bottom
panel) that the short transients due to cosmic hits on the array have
a timescale longer than the chopping time. This allows the
correction of these effects using the ``chop-node'' mode.

We can classify three separate types of transients:
\begin{itemize}
  \item continuum--jump transients: transients due to a sudden change
    of the incident flux from one almost constant level to another
    almost constant level;
  \item cosmic ray transients: transients induced by cosmic ray hits
    which produce glitches, followed by a response variation;
  \item scan dependent transients: transients along a wavelength scan
    produced by rapid variations of the (dominant) telescope
    background and continuum (in the case of a bright target).
\end{itemize}

It is worth mentioning that the Standard Product Generation (SPG)
pipeline, which populates the HSA, does not use any type of transient
correction. So the products available in the archive for the
``frequency switched'' and ``unchopped-mode'' observations contain many
transient effects that could adversely affect science goals. However,
from HIPE 14 onward, users have the option of processing their data
with  special scripts designed specifically to correct transients. The
current paper describes how these corrections are made.
In the following we describe in detail the different types of transients.

\subsection{Continuum-jump transients}

A sudden change in the illumination of a detector pixel passing from
one flux level to another, can induce a major transient in the signal.
This occurs typically at the beginning of each observation just after
the completion of the calibration block. Because of the difference
between the flux from internal calibrators and the telescope
background, a transient is usually visible during the entire
observation.

Another common form of this kind of transients is when a change of
band, or change of wavelength occurs while observing a source. For
example, the user may have requested two different lines occurring at
different wavelengths.  When the grating is commanded to access a
different part of the spectrum, or change to a completely different
region of the spectrum, a jump in the signal usually occurs because of
the changing emission spectrum of the telescope background at
different wavelengths (see Section~\ref{sec:transients}). If the
source is very bright, differences in continuum level from the source
can also induce a transient.

\begin{figure}
\includegraphics[width=1\columnwidth]{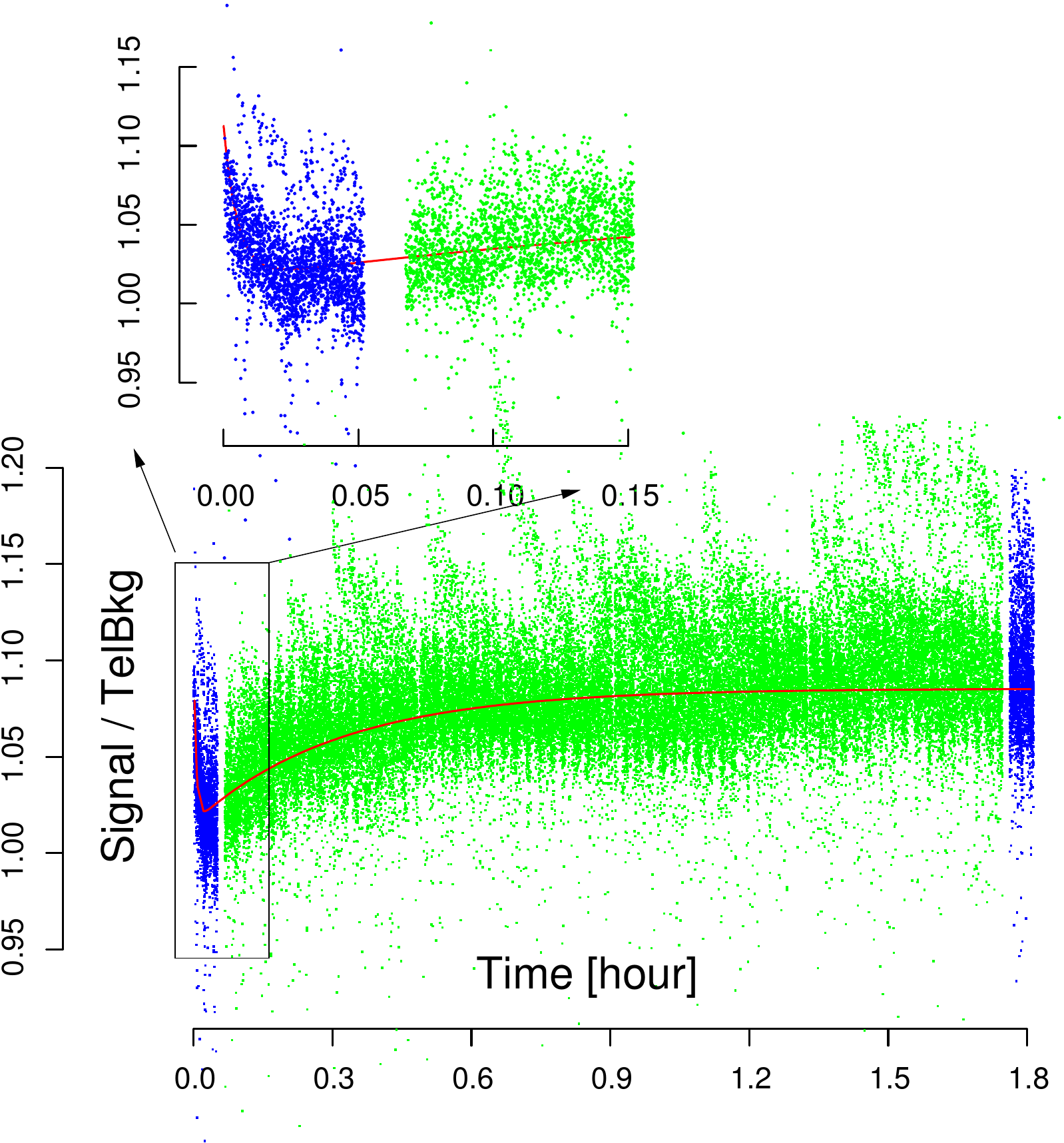}
\caption{ When the signals from a spaxel are normalized to the
  expected telescope background a long-term transient after the
  calibration block becomes clear. In this example (an observation of
  M81, ObsID 1342269535) an OFF position is observed at the beginning
  and at the end of the observation (blue dots), while a 2x2 raster
  observation is performed in the middle (green dots). Since the
  continuum emission of the object is negligible with respect to the
  telescope background, it is possible to fit the general long-term
  transient (red line) and correct the signal. Note that many huge
  transients appear in individual spectral pixels because of cosmic
  ray hits on the detectors.  If this correction is not made, and an
  average background is subtracted from the signal, the final image
  will present an artificial gradient and the flux will be negative in
  some regions. {\it Top inset panel:} a close-up of the initial part of
  the signal just after the calibration block, showing the strong
  transient. The periodic variation in the signal is a left-over from
  the imperfect telescope background estimate used for
  normalization. The empty spaces between points occur during
  telescope slewing.
  \label{fig:ltt} }
\end{figure}

If the source continuum is negligible with respect to the telescope
background, it is possible to normalize the signal during the whole
observation to the telescope background which has been previously
calibrated.  The transient due to the jump in continuum between the
calibration block and the target observation appears clearly, see
Figure~\ref{fig:ltt}.  It is possible, in this case, to fit the
behaviour with a model and subtract it from the observation. It is
interesting to note that, without this correction, an OFF observation
can have a measured flux greater than an ON observation if it happens
to be performed during a period of response stabilization. This effect
is not taken into account by the SPG pipeline, so that many
observations in the archive have an artificial and confusing negative
background.

Some observations that used the unchopped mode contained requests for
more than one band in a single AOR. For extended sources, a typical
raster observation allows one to cover an extended region by moving
the telescope to cover a grid of positions. However, for efficiency
reasons, the AOR was designed to cycle through all the requested bands
at each raster position, before moving the telescope to the next
position in the raster sequence.  An unexpected side effect of this
strategy was the generation of transients at each band change.  It is
not possible to remove these jump--transients in a clean way in case
of such multiple-band observations.

\begin{figure}
  \includegraphics[width=\columnwidth]{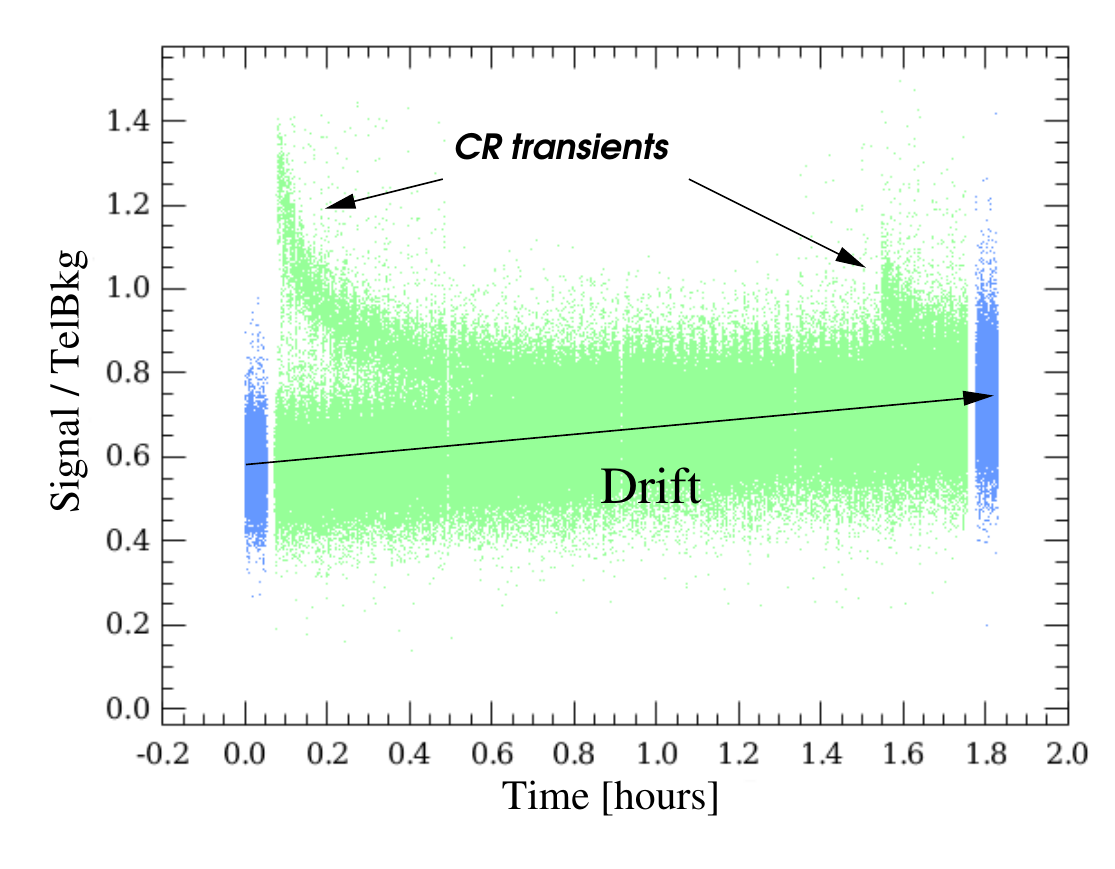}
  \caption{ Long-term transient for the pixels of module 16 in the
    blue channel of the obsid 1342246963.  One of the pixels shows
    also a transient due to a cosmic ray hit.  There is a 30\%
    variation between the two ends of the observation. Blue and green
    dots refer to off-target and on-target parts of the observation,
    respectively.}
  \label{fig:bluedrift}
\end{figure}

There is an important lesson to be learned from this experience with
PACS.  In retrospect, it would have been better to observe the
complete raster in a single band, before cycling to the next
band. This would have minimized the effects of band-induced
jump-transient.  We advise that any future mission where transient
could be an issue, should avoid as much as possible sudden changes in
flux during the observation.  This is probably the case of FIFI-LS
onboard SOFIA, which is essentially a clone of the PACS spectrograph.
Unfortunately, when the observational mode was introduced for {\it
  Herschel}, the reduction pipeline was not fully developed, and it
was very difficult to evaluate all the effects. Since the response was
known to stabilize relatively fast, it was assumed that the effect of
a band change was minimal, compared with the advantage of being more
time efficient. Experience later showed that jump-transient transient
caused by band changes appear to limit the quality of the observation,
even with the most advanced data reduction. Indeed, we are able to see
jump-transients even in the blue channel, which has the fastest
response stabilization compared with the red (see
Figure~\ref{fig:bluedrift}).

\subsection{Cosmic ray transients}

It is well known that for Ge:Ga detectors, energetic cosmic rays
usually produce glitches in the signal, followed by response
variations. Depending on the energy of the cosmic ray, the glitch can
be followed by a tail or the variation can be more complicated
(lowering temporarily the response). A similar behavior was noted
in the past for pixels in the ISOCAM array on the Infrared Space
Observatory \citep[see, for example, ][]{2001MNRAS.325.1173L}. We will
show that the response variation can be described with a combination
of exponential functions (see Section~\ref{sec:transmodel}).  The main challenge to
correcting and masking the damage in the signal from cosmic ray hits
is to select the most significant events, and then find the starting
time which marks the beginning of the discontinuity in the signal.

\subsection{Scan dependent transients}
\label{sec:scandeptrans}

Finally, in the case of observations spanning an extended wavelength
range, rapid variation of the continuum during the wavelength scan
produced transients in the signal (see
Section~\ref{sec:scantransients} for an example). In some
observations, this last effect is responsible for different apparent
fluxes in the source spectrum for upward and downward scans over the
same wavelength range. Luckily, many PACS observations using the
unchopped mode targeted only one line with a short range scan.  For
these cases, scan-dependent transients had negligible effects.  It was
found to be important only for observations which scanned over
extended wavelength ranges.  As a result, two different interactive
pipeline scripts (accessible within HIPE as so--called {\it ipipe}
scripts) have been written to treat the ``unchopped line'' and
``unchopped range'' cases separately.

\section{Detecting and modeling transients}
\label{sec:transmodel}

To detect, model, and correct transients in the signal, one has to
decouple the signal and the effect of the transients. The best way to
proceed is to obtain an estimate of the response by normalizing the
signal to the expected spectrum.  As we will see in the following
sections, if the source emission is negligible with respect to the
telescope background, the ideal way to proceed is to normalize the
signal to the expected telescope background.  If the source emission
is not negligible, one has to have a better guess of the signal. One
possibility is to assume that the detector eventually stabilizes after
a jump in the continuum, so that one can use the last scans near the
end of the observing sequence to estimate the ``true'' continuum level.
This asymptotic level can then be used to normalize the entire signal
  to reveal the jump--transient. Once the transient is revealed in this
  manner, it can be fitted and removed from the signal.

Once the transients due to expected continuum-jumps are corrected, it
is possible to obtain a first estimate of the final spectrum,
combining all data from all independent up and down scans, and
measurements made with all 16 pixels combined together.  Once this
average spectrum is created, it can be used to revisit the individual
pixel responses by normalizing their signal to this ``first guess''
spectrum.  In this way, each individual detector pixel response to can
be studied for the effects of cosmic ray transients, and these CR
transients can be either corrected or masked.  In the following, we
describe the model used to fit the response as a function of time and
the algorithms applied to the signal to mitigate the effect of the
transients.

\begin{figure}
\includegraphics[width=1\columnwidth]{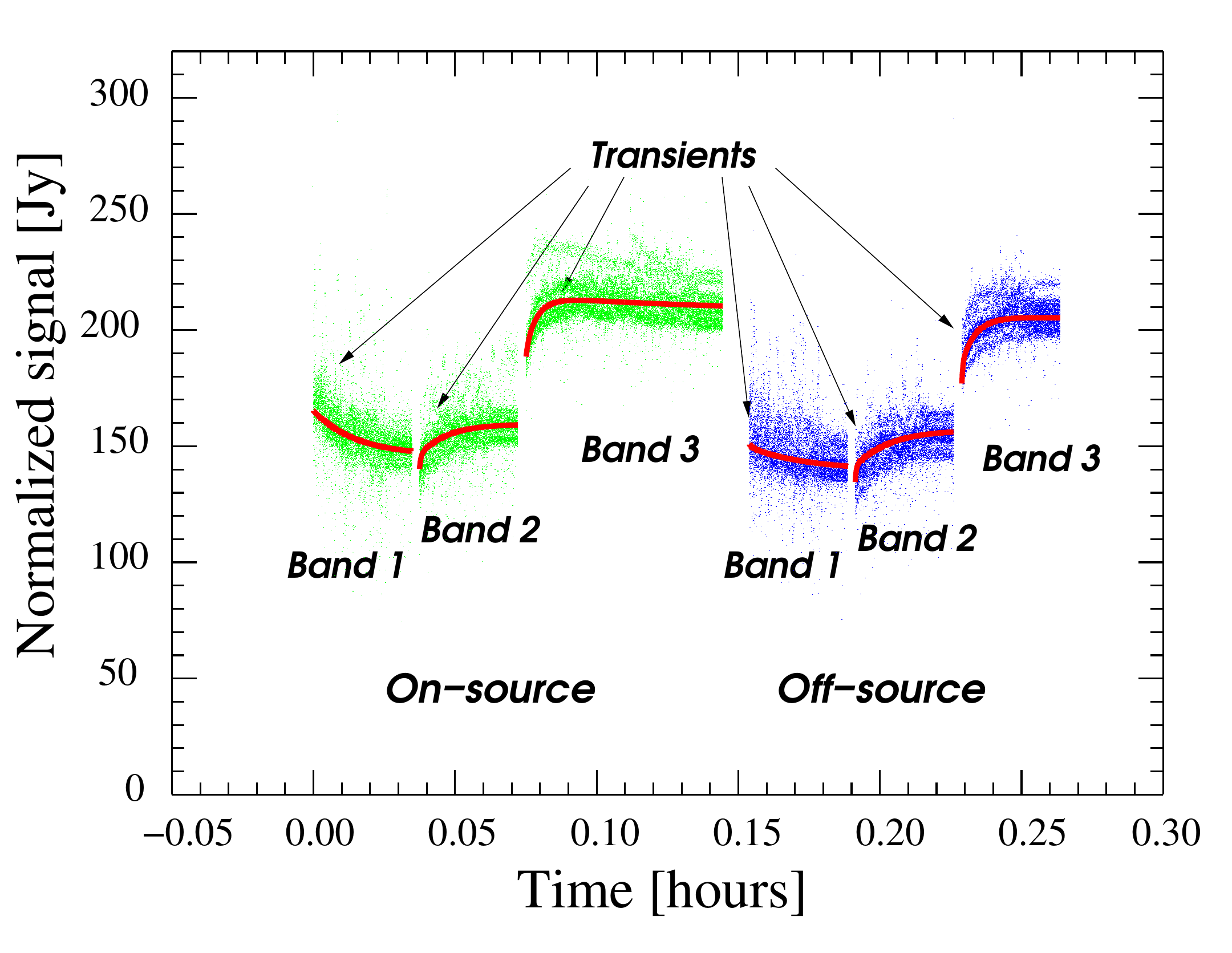}
\caption{ The central spaxel of an observation of NGC6543 (ObsID
  1342215704) with three lines observed. For each observation block,
  the signal in each wavelength scan is divided by the last scan and normalized to its asymptotic median flux.
  The red line shows the transient model fitted.
  The first block comes after the calibration block (not shown)
  whose flux is higher than the continuum during this part of the
  observation. The three lines are observed first on the target (green
  dots) and then on the OFF position (blue dots). The gaps in these
  science data correspond to the time spent moving the grating, and
  slewing to the OFF position. Also in this case some pixels appear to
  behave differently because of the impact of cosmic rays.  }
\label{fig:lttblock}
\end{figure}

\subsection{Model}

The response variations can be described in general using a combination of
exponential functions. We found that a combination of three exponential
functions closely describe the several different cases of transients
observed.

\begin{equation}
  f(t) = a + a_{short} e^{-t/\tau_{short}}+ a_{medium} e^{-t/\tau_{medium}}+ a_{long} e^{-t/\tau_{long}}
\end{equation}

The parameter $a$ corresponds to the median level of the normalized
signal.  In the case that the response is obtained by normalizing the
signal to the last scans, $a$ is fixed to unity by construction. The
three exponential functions have very different timescales. A first
function accounts for the fast variation of the signal just after the
flux change or a cosmic hit. The timescale is between 1 and 10
seconds. The second term has an intermediate timescale (between 10 and
80 seconds). Finally, the third term accounts for the long-term
variation of the signal and has the longest timescale (between 200 and
1200 seconds).  In the case of cosmic ray induced transients, the
$a_{short}$ coefficient is always positive, whereas for the
continuum-jump transients $a_{short}$ can be either positive or
negative depending on the direction of the jump.  The parameter space
considered for the fitting in the two cases is similar.

In the case of cosmic rays, the time constant $\tau_{short}$ is kept
lower than one second. For the continuum-jump transient, the parameter
space for $\tau_{short}$ considered is between 1 and 10 seconds.  For
the other two time constants, the parameter boundaries are
$10<\tau_{medium}<80$ seconds and $\tau_{long}>100$ seconds.

\subsection{Correcting continuum-jump transients}
\label{sec:transients}

If the continuum of the source is negligible with respect to the
telescope background, it is possible to see the long-term transient
caused by the calibration block all along the observation by
normalizing the signal to the expected telescope background (see
Figure~\ref{fig:ltt}).  If this effect is not corrected and an average
background is subtracted from the signal, there will be parts of the
image with a signal lower than the average background which will have
negative fluxes at the end of the data reduction.  Even worse, in
mapping observations, each raster position taken at a different phase
of the transient will have a different continuum level resulting in an
artificial gradient in the final image (see later in
Figure~\ref{fig:gradient}).

Unfortunately, this correction depends on the knowledge of the
telescope background emission. The current model available through
HIPE is fairly accurate for the red array, while the situation is
slightly worse for the blue array.

There are several situations where the simple correction above will
not work. In the case of a very bright continuum, the normalization of
the signal by the telescope background will not be appropriate, and
this will make any correction difficult.  Another case is that of a
multi-band observation taken consecutively. In
Figure~\ref{fig:lttblock} we show an example of an observation taken
in 3 different bands. At the beginning of each observational block
(observation of one band) there is a clear transient introduced by the
sudden variation of the continuum level.

To correct the transients in both of these cases, we can normalize the
signal to the flux detected during the last wavelength scan of each
observational block. We make the assumption that the detector response
has stabilized by this time. This assumption works well in the case of
long observations ($\sim$1 hr). In the case of shorter observations, a
little bias will remain in the data since the asymptotic level has not
yet been reached.

\subsection{Correcting transients caused by cosmic rays}

The most disruptive transients are caused by the impact of cosmic rays
on the detectors.  Since they happen at random times with different
energies, they cause unpredictable changes in the baseline of the data
which can last from several minutes up to one hour.  The main
difficulty in fitting these transients with our model is to
disentangle the real signal from the variation in response caused by
the cosmic ray impact, and to detect the time when the impact
occurs. Cosmic ray events are so frequent that, in practice, we must
set a threshold above which we attempt to correct the signal.

\citet{2012A&A...548A..91L} developed a procedure to identify features
associated with transients using a non-parametric method
(multi-resolution transform of the signal and identification of
patterns on a typical scale). The features found were then fitted with
an exponential, and subtracted from the signal. This algorithm is
identical to the one developed for the treatment of ISOCAM data
\citep{1999A&AS..138..365S}.

The approach we described in our paper is different in both the way we
identify the cosmic ray events, and how remove them.  Our methods are
similar to those developed by \citet{2001MNRAS.325.1173L} for the
treatment of ISOCAM data, although the algorithms presented here
differs in several significant details. First we obtain an
approximation to the final spectrum cladding all the individual pixels
that contribute to the scan (the ``Guess'' spectrum). Next, the signal
for each individual pixel is normalized by this spectrum to obtain an
estimate of the response. At this point, the discontinuities in the
response are identified, and only the most significant are
selected. We fit the model (equation 5) to the response in the
interval between two consecutive discontinuities.  A correction is
then applied to the original signal to attempt to remove the effect of
the cosmic rays. In the following, we describe in detail each one of
these steps.

\begin{figure}
\begin{center}
  \vbox{
\includegraphics[width=0.6\columnwidth]{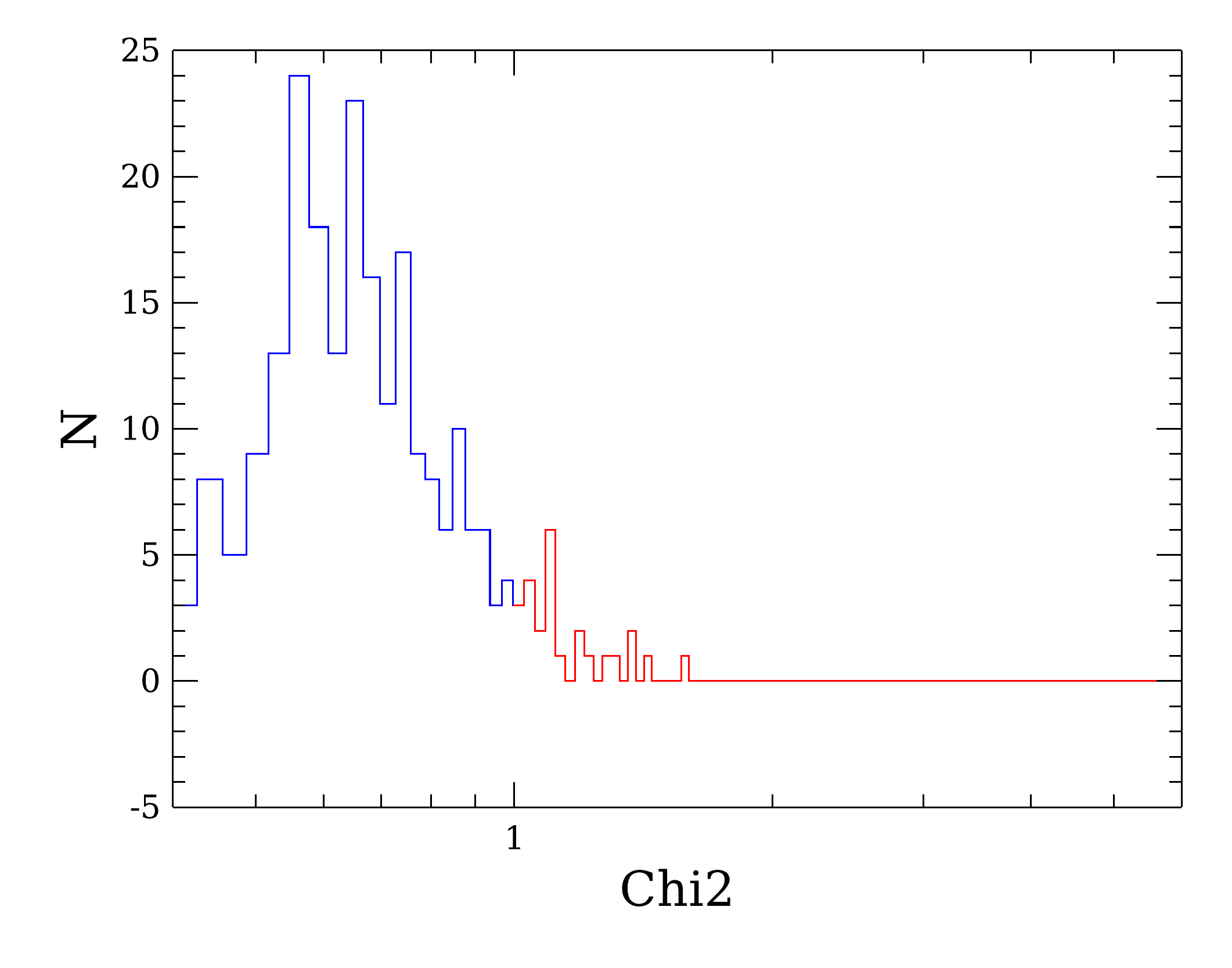}
\includegraphics[width=1\columnwidth]{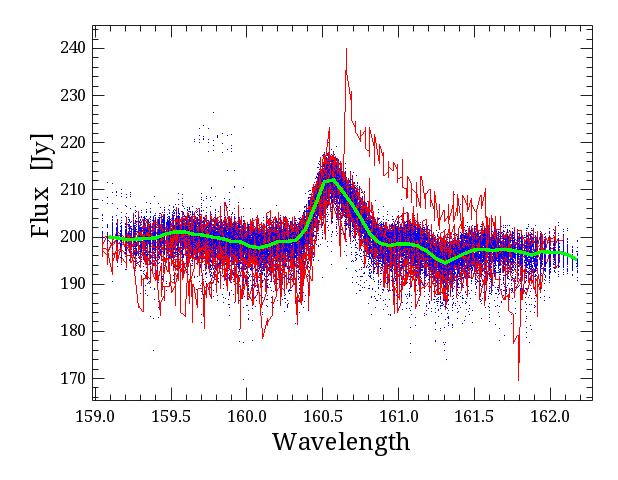}
  }
  \end{center}
\caption{An example of how we estimate the ``Guess'' spectrum for the C+
  line of Arp~220 (obsID 1342202119). We compute the histogram of the
  normalized $\chi^2$ of each individual spectral scan with respect to
  the median spectrum (top panel). The blue line shows the well
  behaved scans and the red line shows those to be rejected. The Guess
  spectrum (bottom panel; green line) is the result of co-adding the
  well behaved scans (blue lines). The scans with $\chi^2$ deviating
  more than 3$\sigma$ from the median value are discarded (red lines).
}
\label{fig:guess}
\end{figure}

\begin{figure}
\begin{center}
 \includegraphics[width=0.9\columnwidth]{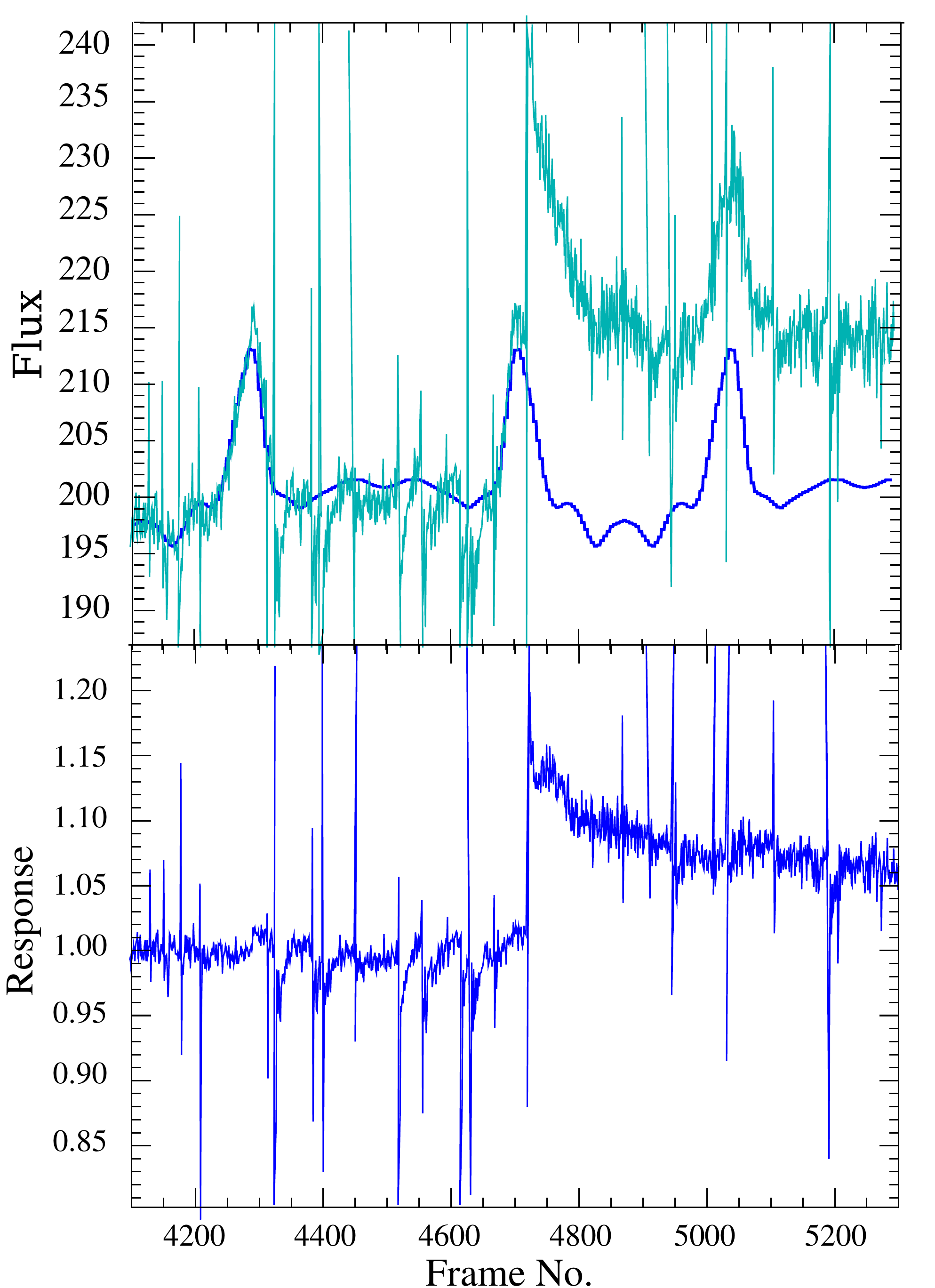}  
  \end{center}
\caption{{\it Top panel:} A comparison of the observed signal of a single
  spectral pixel (cyan) with the ``Guess'' signal (blue; see text). The
  signal is plotted against the Frame Number, which is a function of
  time during the scan.  We show three scans in the figure (two up and
  one down) which captures the C+ line shown in the previous figure.
  Just after the frame number 4720 a strong cosmic ray hits the
  detector producing a sudden change of the response of the detector.
  {\it Bottom panel:} The ratio of the two curves, i.e. the response of the
  pixel as a function of frame number. Many transients are clearly
  visible, in particular the huge transient after frame number 4720.}
\label{fig:expected}
\end{figure}

\begin{figure}
\begin{center}
  \includegraphics[width=0.9\columnwidth]{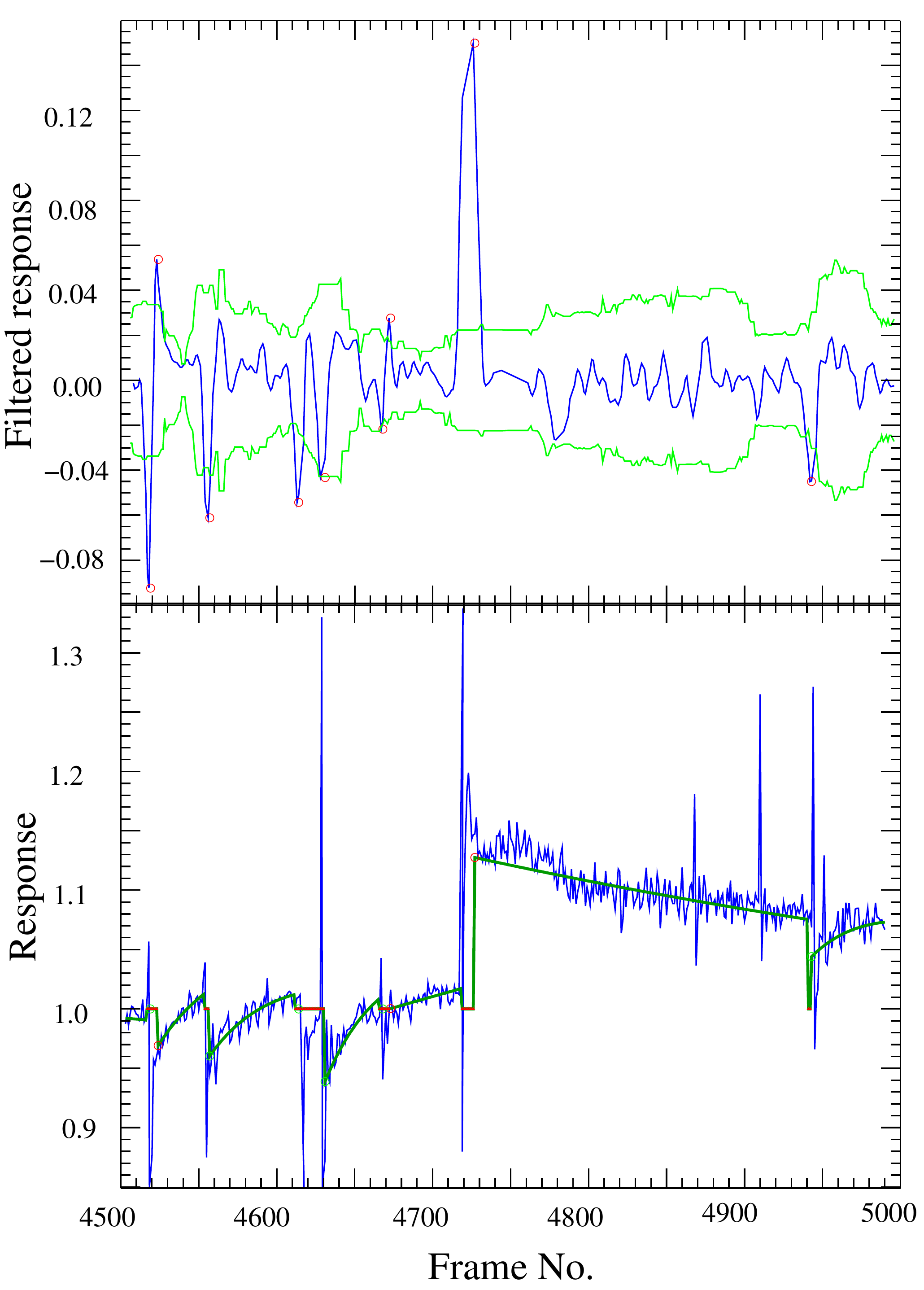}
\end{center}
\caption{On the top panel, part of the response convolved with the
  first derivative of the Gaussian with dispersion equal to two frame
  units. This width gives the best sensitivity to sudden changes of
  the response caused by cosmic ray hits on the detectors. The green
  lines mark the region of $\pm 5$ times the local dispersion and the
  red circles marks the starting points of transients.  In the bottom
  panel, the responsivity for the same block of frames with each
  interval between two consecutive discontinuities in the response
  fitted with our transient model (dark green). The red points mark
  the frame masked because the interval between two consecutive
  discontinuities is too short (less than 10 frames).  }
\label{fig:filter}
\end{figure}

\subsubsection {Estimating the ``Guess'' spectrum}

We assume that the continuum jump transients have been corrected to
this point.  We need to identify the scans affected by severe
transients, and discard them to obtain a good first guess of the
spectrum.  This step is particularly important in the case of low
redundancy, since even a few bad scans can compromise the coadded
spectrum.  The procedure to obtain a guess spectrum is iterative. The
first step consists of computing a median spectrum by coadding all the
wavelength scans for the spectral pixels which contribute to a given
spaxel.  The spectrum is obtained by computing the median of all the
valid fluxes for each bin of a wavelength grid: the median
spectrum. Scans with very deviating values compared with the median
spectrum are rejected.  The scans with $\chi^2$ values which deviate
more than 3$\sigma$ from the median of the $\chi^2$ distribution are
discarded and a new median spectrum is computed (see top panel of
Figure~\ref{fig:guess}).  To obtain a robust estimate of the
dispersion of the distribution, we make use of the median absolute
deviation (or MAD).  The procedure is iterated three times to obtain a
clean median spectrum which serves as ``Guess''
spectrum. Figure~\ref{fig:guess} shows visually how the process works
in the case of an observation of the C+ line of Arp~220. The procedure
works because of the high redundancy of the PACS data, which has a
minimum of four up-and-down scans for each of the 16 spectral pixels.

\subsubsection{Finding significant discontinuities}

By normalizing the signal of each pixel by the expected flux of the
Guess spectrum, we obtain an estimate of the response as a function of
time for each pixel. This function can be used to study the effect of
cosmic rays on the signal by selecting the most important ones,
fitting the transients after them, and eventually masking part of the
signal excessively damaged by the effect of cosmic ray impacts on the
detector. In Figure~\ref{fig:expected} we show an example using the
M82 C+ spectrum. Here the Guess signal is compared with the signal
from a single pixel, allowing the discontinuities to be easily seen.

To correctly fit the transients, one has to identify the starting
point, i.e. the moment at which the discontinuity in flux appears in
the signal.  This is a classical problem of signal theory and it has
been shown \citep{Canny1986} that the optimal filter to find
discontinuities in a mono-dimensional signal is the first derivative
of a Gaussian.  The dispersion of the Gaussian was found empirically
to be two frame units (equivalent to 0.25 s).  If we convolve the
response with this filter, spikes appear just after cosmic rays impact
the detector.  As shown in Figure~\ref{fig:filter}, we can select the
most important ones by comparing their intensity to the local noise in
our response estimate. In our algorithm we make a cut of 5~$\sigma$ to
select the significant events.

\begin{figure}
  \begin{center}
    \includegraphics[width=0.9\columnwidth]{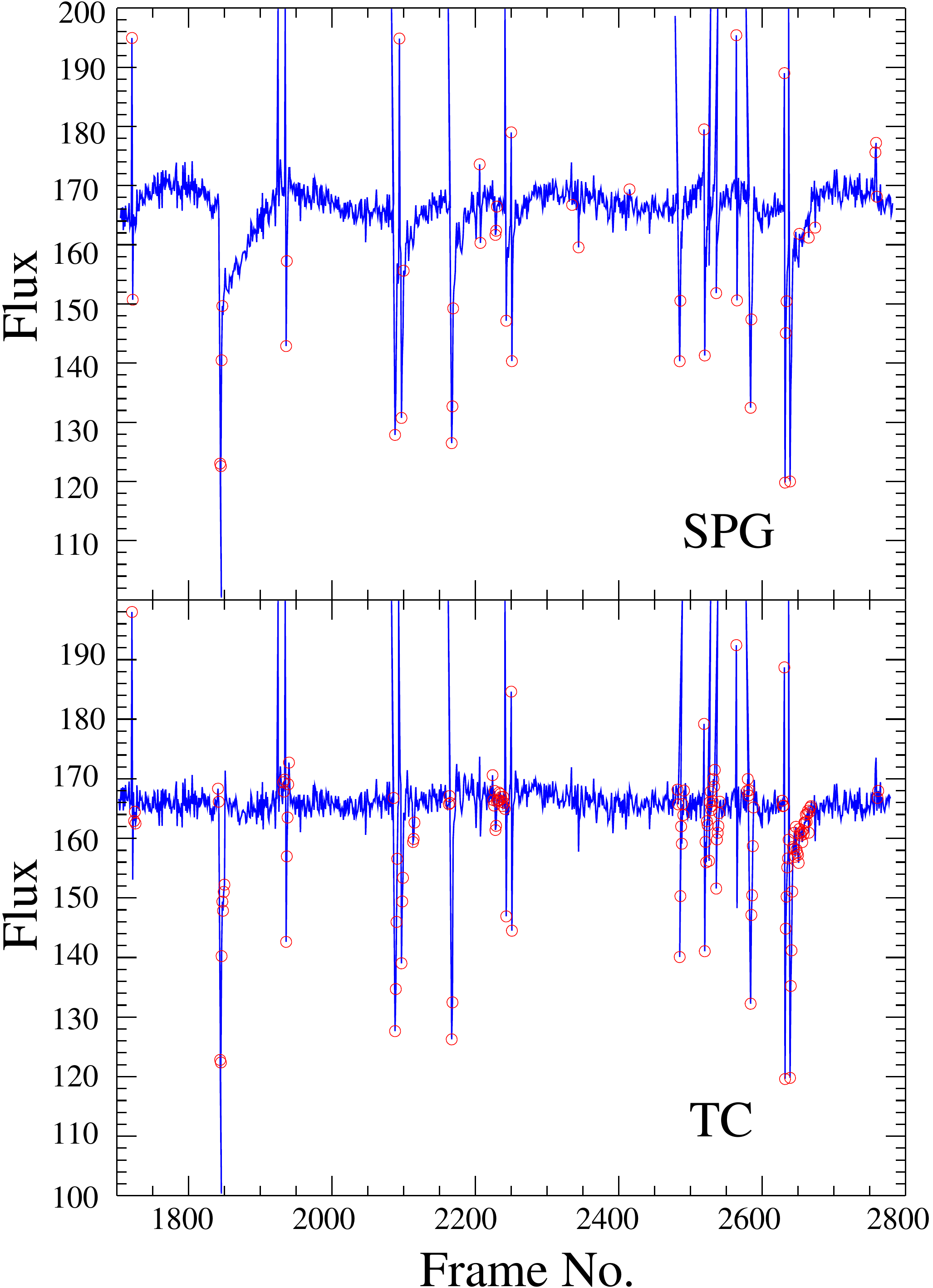}
    \end{center}
\caption{Comparison between masking in the SPG pipeline (top) and
  transient-correction pipeline (bottom). Signal is blue and masked
  frames are represented as red circles. The SPG pipeline masks a
  few deviant frames as GLITCHES, while the transient-correction pipeline
  masks frames either as GLITCHES or as UNCORRECTED. As a result, the signal
  from the SPG pipeline will contain unmasked residuals of big transients,
  while the transient-correction pipeline will either correct the transients
  or masked them completely. This leads to a big reduction of noise in the
  final reduction with respect to the product from the SPG pipeline.
}
\label{fig:residuals}
\end{figure}

\subsubsection{Masking and computing the response}
At this point we break the signal in intervals between two consecutive
discontinuities, and fit the model (equation 5) in each one of these
intervals. If the cosmic rays occur too close together (less than 10
frame units) we simply mask that part of the signal timeline.  The
resulting model response and masked regions are then applied to the
original signal to remove the effect of the transients.  This
automatically corrects the signal for variations in the response after
cosmic ray impacts, as well as correcting for large offsets in
individual scans relative to the median signal.

One main difference with the reduction pipeline for chopped
spectroscopy is that, at this point, there is no need for the
application of a spectral flat field correction.  In the standard
pipeline, this is a separate task that takes into account small
variations of the response in the different spectral pixels. Our
modeling is able to automatically rescale all the pixels relative to
the same spatial module in each observational block without any further operation.

The net effect of all our corrections is to significantly decrease the
noise in the final coadded spectrum compared with the SPG pipeline.
Although the SPG pipeline attempts to mask larger transients,
important residuals remain (see Figure~\ref{fig:residuals}).

\begin{figure}
  \includegraphics[width=\columnwidth]{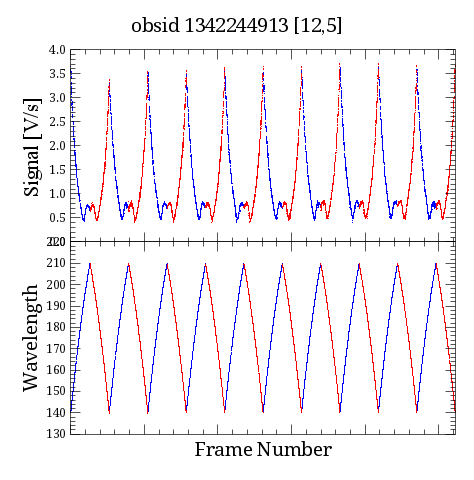}
  \caption{ Signal variations as a function of the
    wavelength scan. To see the effect of the flux variations on the
    measured signal, we can compare the signal measured during the
    up-scans (blue parts) with the one measured during the down-scans
    (red parts).  }
  \label{fig:updown}
\end{figure}
\begin{figure}
  \begin{center}
    \includegraphics[width=0.8\columnwidth]{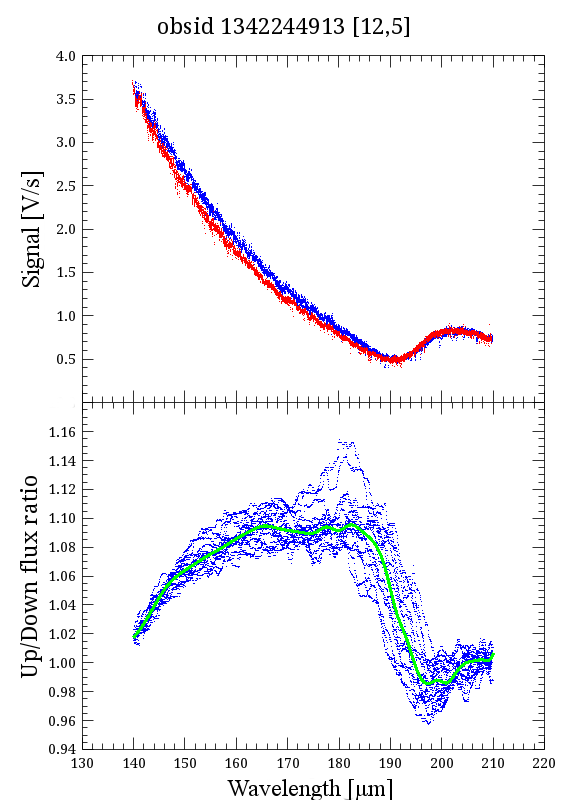}
  \end{center}
  \caption{ {\it Top panel:} flux in the up-scan (blue) compared to the flux
    in the down-scan (red). When the incident flux is decreasing (blue
    line between 140 and 195~$\mu m$) the measured flux is higher than
    that measured when the flux is increasing (red line in the same
    wavelength region). The situation is inverted between 195 and
    205~$\mu m$, while the two fluxes are very similar in the
    wavelength region beyond 205~$\mu m$ where there is no big
    variation in flux. {\it Bottom panel:} ratio between up- and down-scan
    measured fluxes for all the 16 pixel in a space module. The
    behaviour is very similar.  }
  \label{fig:updownratio}
\end{figure}

\subsection{Correcting scan dependent transients}
\label{sec:scantransients}

For many of the PACS spectra, only a small wavelength range is
observed. In this case we can consider that the continuum is
essentially constant during the observation. So, during a wavelength
scan there is almost no variation in the incident flux which could
cause transient effects on the signal.  This is not the case in
observations of extended wavelength ranges, where the continuum can
vary substantially during the scan. This leads to transient behaviour
which becomes apparent when comparing the scans in two opposite
directions.  In Figure~\ref{fig:updown}, we show that during up and
down scans, the incident flux can change by, in this example, a factor
of 7.  This leads to a large transient.

In Figure~\ref{fig:updownratio} (top panel) the ratio of the two
fluxes shows clearly the effect. Here we consider only one scan for a
given spaxel.  For the up-scan (blue points), we see the flux
decreases as we scan to longer wavelengths. However, when we scan back
with a down-scan, (red points) there is a systematically lower signal
compared with the up scan. This is due to transient behavior caused by
the rapidly changing flux during the scans.  The effect is clearly
shown in the ratio of the up and down scans (bottom panel of
Figure~\ref{fig:updownratio}).  When the flux is changing very rapidly
(between~145--190~$\mu$m), the up-scan flux is 10\% higher than the
down-scan flux. However, from~190--210~$\mu$m, where the flux change is
much smaller (only a factor of two), the effect of the transient is
also smaller (less than 2\%).

\begin{figure}[th!]
  \begin{center}
    \includegraphics[width=\columnwidth]{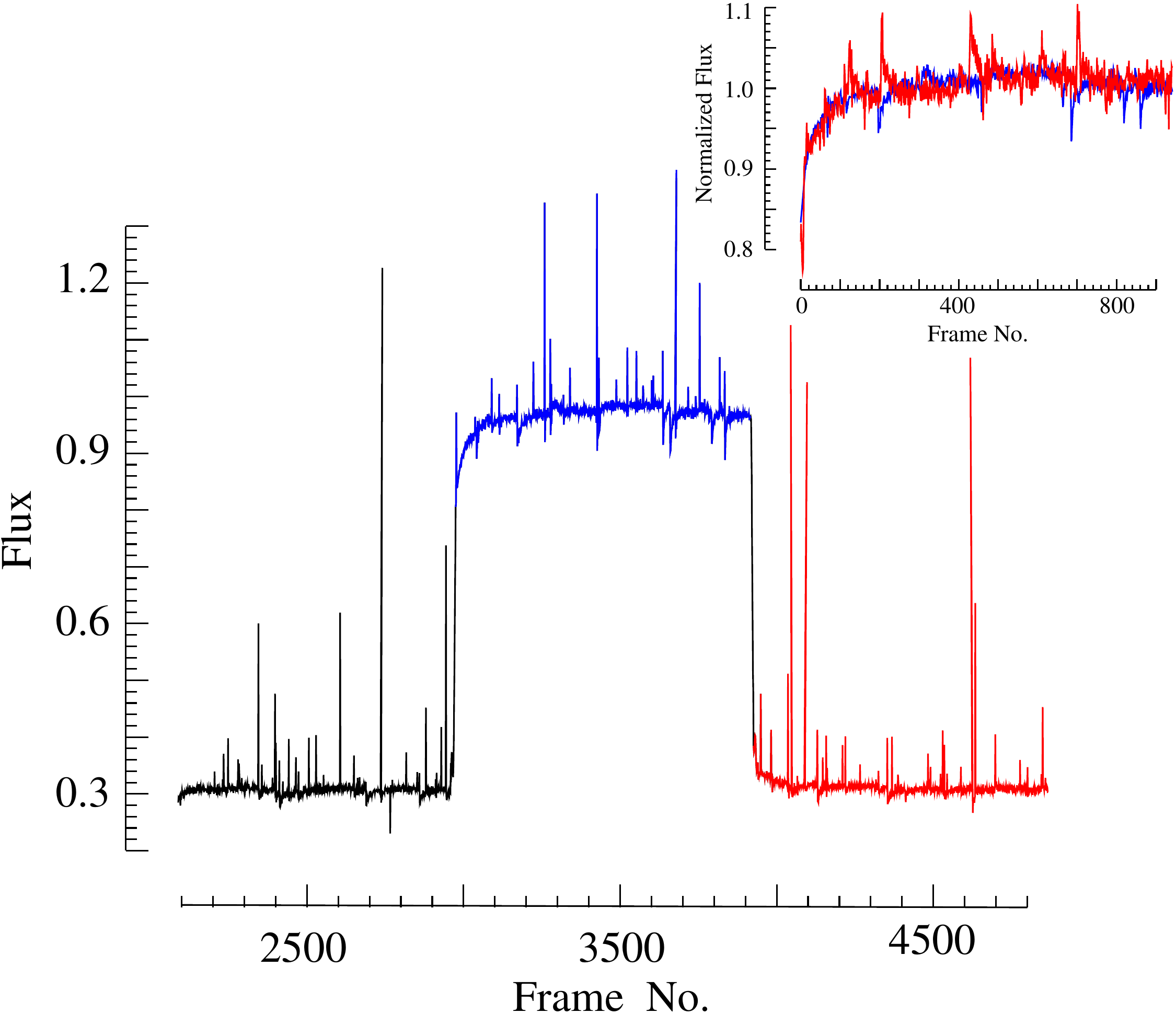}
  \end{center}
  \caption{ Transient effects on the measured signal when changing
    between two grating positions performed during an engineering
    observation during OD 27 (see text).  In the inset, comparison of
    the upward (blue) and downward (red) transients.  The downward
    transient has been flipped to compare the two behaviours.  For a
    regime of flux changes similar to the one found in the range
    observations, the two transients follow essentially the same
    behaviour. This justifies the choice of getting the median signal
    between up- and down-scan signals. }
  \label{fig:od27}
\end{figure}

Before we describe in detail how we apply a correction for this kind
of transient, we first need to demonstrate that the transient caused
by a sudden increase in continuum has a similar form to that of a
sudden decrease in continuum. We will use this symmetrical behaviour
of transients to provide a solution to the problem.  To prove this
symmetry, we consider an observation which was especially designed to
induce transient behaviour of the detectors during observational day
27 (obsID 1342178054).  In this observation, the grating was moved to
several different positions to have different levels of flux on the
detectors. After each move, the grating was halted for 4 minutes to
allow stabilization.  We show in Figure~\ref{fig:od27} a part of this
observation to demonstrate the response to sudden jumps in recorded
flux. The figure includes the signal readout just before (black line)
a sudden change in the position of the grating, which then induces an
upward transient and new stabilization (blue line). After a suitable
interval, the grating is changed back to the original position leading
to a downward jump and transition (red line), followed again by a
period of stabilization. To investigate the shape of the upward and
downward transients induced by this change in continuum, we show in
the inset figure, the normalized signal for the upward-- and
downward--going signal, but with the downward signal flipped in sign
to allow a close comparison. If we ignore the cosmic ray transients
which are present, the behavior of the upward--going (blue) and
downward--going (red) transient responses are almost identical.

In the above example, the largest part of the transient response
occurs within a time interval of approximately one minute, and the
example we chose corresponded to a change in flux of about a factor of
three. We recall that in Figure~\ref{fig:updownratio}, the up-and-down
scan induced changes in continuum which were larger than this over the
whole scan (which lasted two minutes), but are about the same order
(roughly a factor of three) in one minute. Therefore, the test of the
hypothesis that the symmetry in the shape of the upward and downward
transient is valid.

If we make the assumption that indeed the upward-- and downward--going
transients have similar behaviors (except for the sign), this suggests
a workable way to correct for the transients in a long scan over a
rapidly changing continuum. Because of the symmetry, we can assume
that the differences between the fluxes in
Figure~\ref{fig:updownratio} is entirely due to transient behavior.
This means that a reasonable solution is to scale the up- and
down-scan values of the flux to the average of the two values at each
wavelength interval. With this simple solution, according to the observation
in Figure~\ref{fig:od27}, the flux recovered is within a few percent of the
asymptotic incident flux. We note that a change of a factor of
three in the continuum in one minute (as in the above example) is an
extreme case. In most observations, the changes in flux throughout a
scan are likely to be much less.

\section{Examples}

In the following section we provide a few examples which show the
improvements introduced by our transient-correction algorithms, as
compared with the standard unchopped-mode HIPE analysis (SPG pipeline,
vers. 14).  We remind that it is important to avoid executing pipeline
scripts blindly without examining the effects of the corrections in
the different steps of the reduction process.

\begin{figure}
  \vbox{
  \hbox{
  \includegraphics[width=.45\columnwidth]{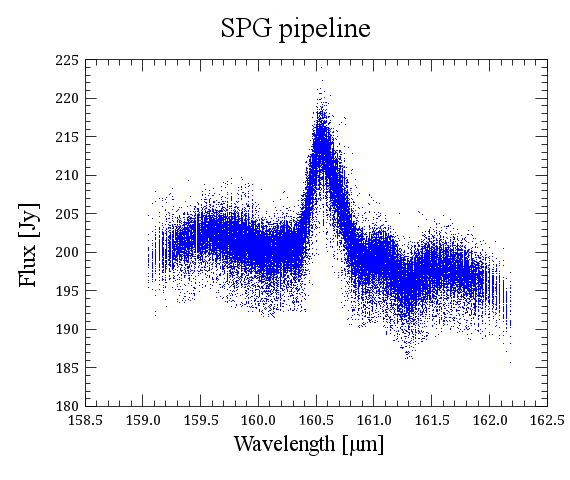}
  \includegraphics[width=.45\columnwidth]{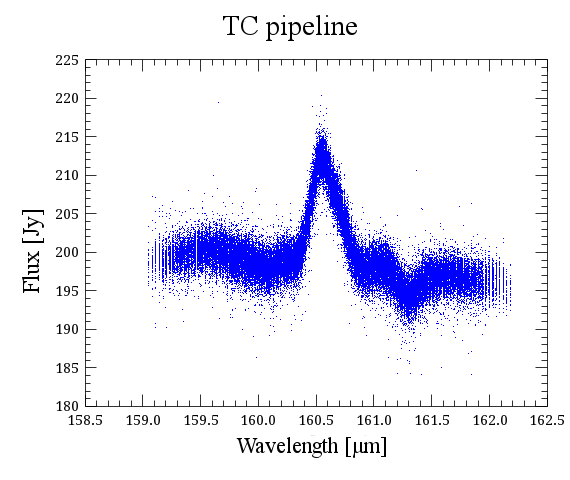}
  }
  \includegraphics[width=.9\columnwidth]{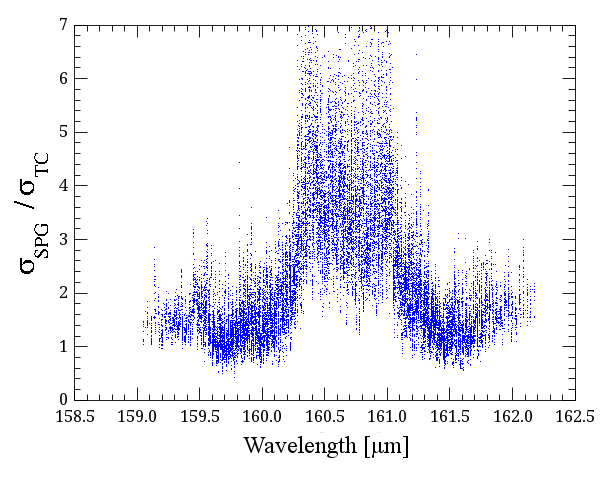}
  }
  \caption{
    {\sl Top }: SPG and transient-correction pipeline reductions of the central pixel of an observation of Arp~220 (obsID 1342202119).
    {\sl Bottom}: ratio of standard deviations of the SPG and the transient-correction pipelines.
  }
  \label{fig:line}
\end{figure}

In this section we also show a few comparisons of observations made
with unchopped and chop-nod mode of the same targets, to show how the
reduction described in this paper gives consistent flux calibration
for the same observed lines.

\subsection{Line}

As an example of the improvement in the reduction of a line unchopped
observation, we show here an observation of Arp~220 (obsID
1342202119).  We show in Figure~\ref{fig:line} (top left panel) the cloud
of flux--points for the central spaxel of the spectrometer array after
passing the raw data through the SPG pipeline.  In the top right panel of
the same figure, the same observation was processed using the
algorithms described in this paper. It is clear that the
points are much less spread in the case of the transient--corrected data compared with
the SPG pipeline.  To evaluate the improvement in a more quantitative
way, we computed for a given flux point, the dispersion in the
surrounding 30 points, and then plot the ratio between the values for
the two pipelines (bottom panel).  The SPG pipeline has a standard deviation
40\% larger in the continuum surrounding the line, as compared with
the transient-corrected data.  On the C+ line itself, the standard deviation
is $3\times$ larger for the same comparison.  We note that the faint
absorption feature at 161.3~$\mu$m is much better defined in the
reduction with transient corrections. The noise of the SPG pipeline is
twice the value of the transient corrected pipeline on this spectral
feature. This demonstrates the power of the new reduction methods.

\subsection{Range}

In this section we compare the reductions of range-scan observations
from the archive (SPG 14) and our pipeline. For this comparison we
used the same version of calibration data used in the archive (version 72).
Among several tests done, we present two representative cases:
the molecular cloud Mon R2 (AORs 1342228456/7) and the galaxy NGC 6303
(AORs 1342214685/6).

\begin{figure}
  \vbox{
    \hbox{
      \includegraphics[width=0.6\columnwidth]{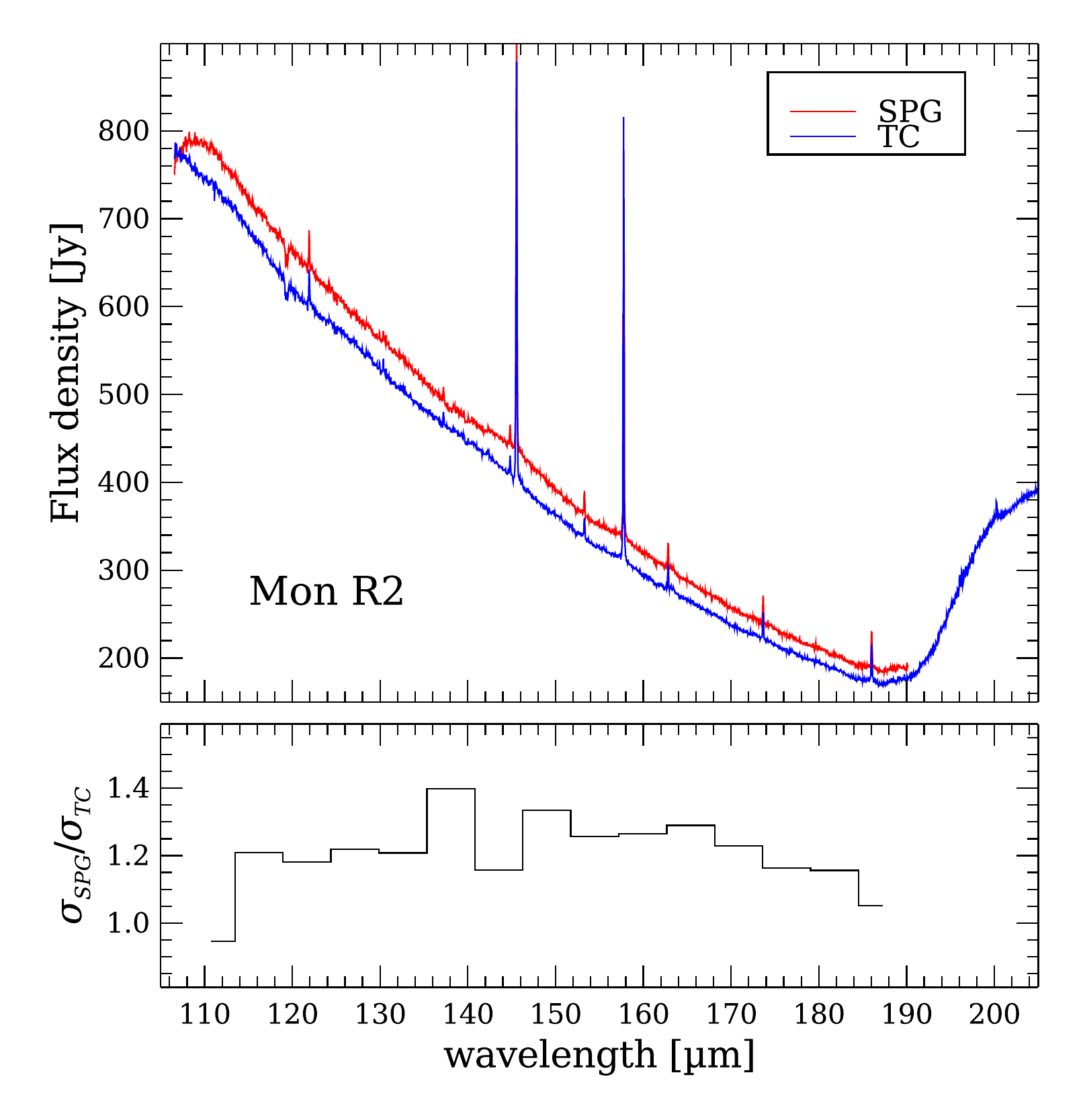}
      \vbox{
        \includegraphics[width=0.4\columnwidth]{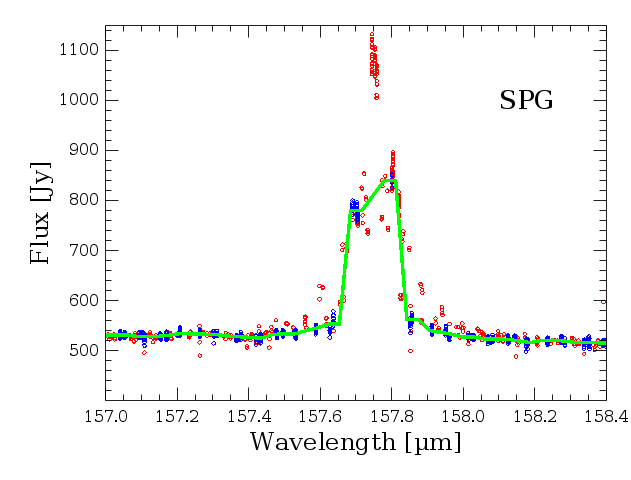}\\
        \includegraphics[width=0.4\columnwidth]{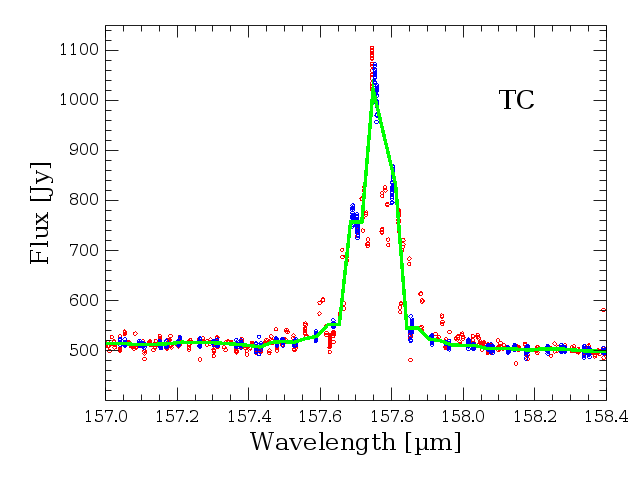}
      }
    }
    \hbox{
      \includegraphics[width=0.6\columnwidth]{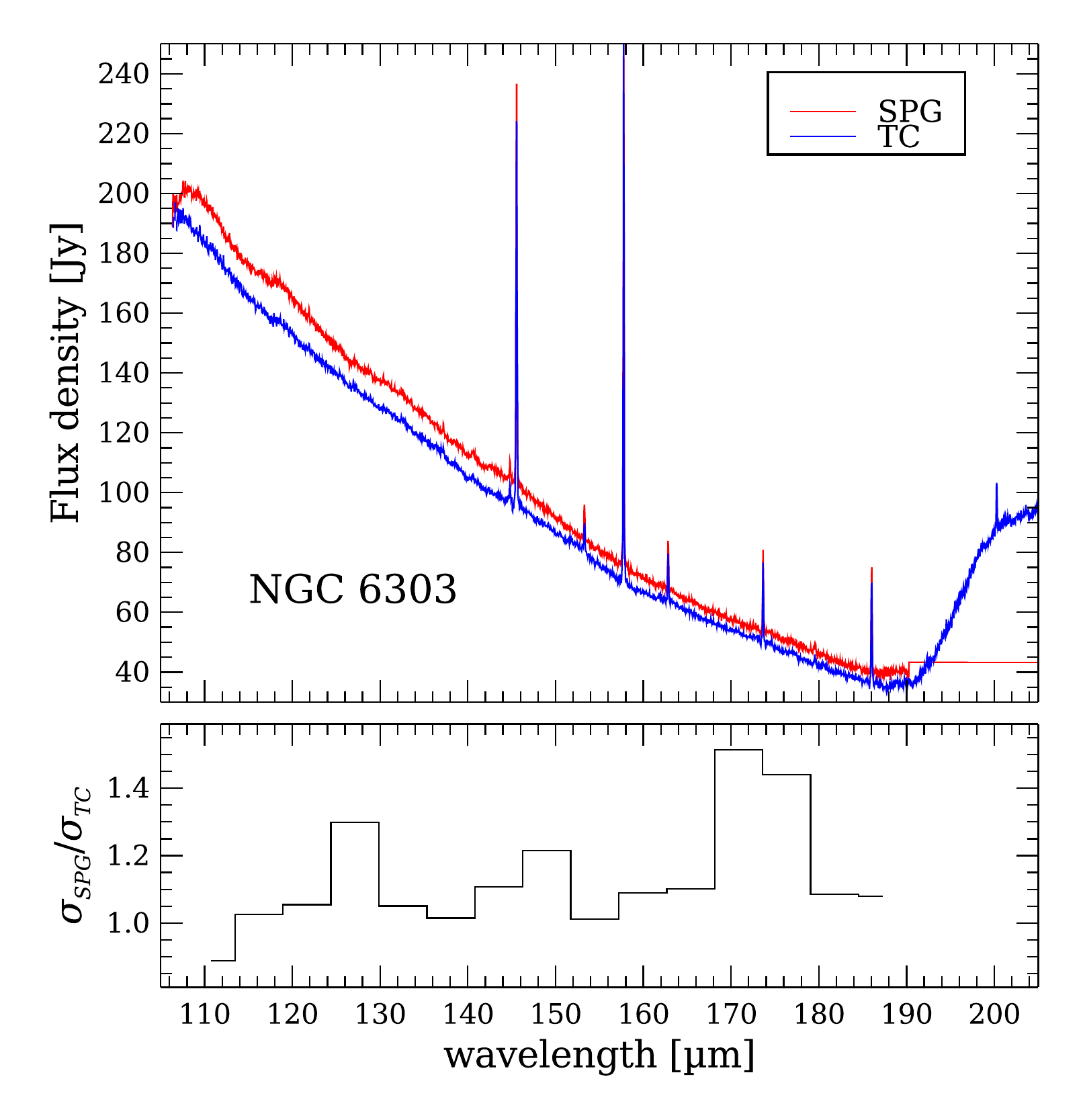}
      \vbox{
        \includegraphics[width=0.4\columnwidth]{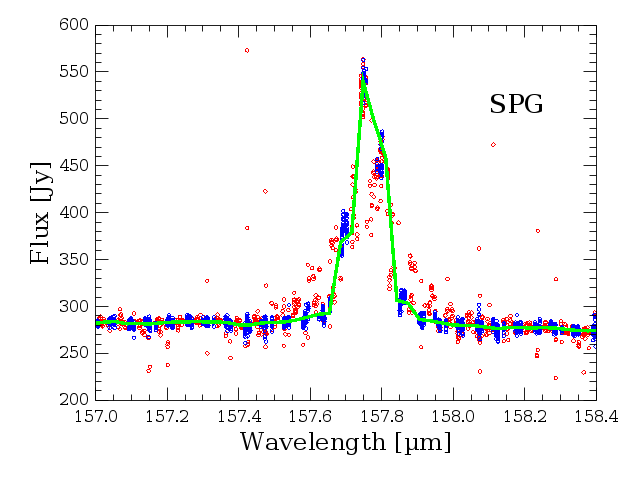}\\
          \includegraphics[width=0.4\columnwidth]{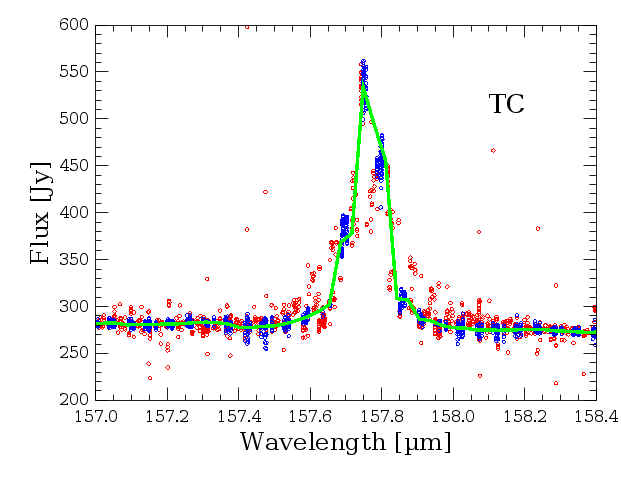}
      }
    }
  }
  \caption{
    {\em Left:} Spectra of Mon R2 and NGC 6303 from the archive products (SPG, red) and reduced with our pipeline (TC, blue).
    {\em Right:} [CII] line as recovered by the SPG and TC pipelines for the two sources before background subtraction. Masked frames are shown as red points.
  }
  \label{fig:rangecomp}
\end{figure}

The latest archival products of this mode processed with HIPE 14 show
a significant improvement with respect to those processed with HIPE 13
as flat corrections were used. As visible in
Figure~\ref{fig:rangecomp}, the current archival products are very
similar to our pipeline reductions. In particular, the shape and
intensity of lines are very close. There are however differences in
the absolute level of the continuum, broad features of the continuum,
noise level, and shape of very bright lines.
As already explained,
range observations require two different AORs to observe the target
and an OFF position. Since the two observations are not connected and
could be taken in different times, it is possible that subtracting the
resulting spectra will not yield a correct continuum level. When
applying a correction for transients, this can lead to differences in
the level of the continuum since one AOR can suffer from a stronger
transient than the other one.
Transients are also responsible for the
unphysical turn-around of the SPG spectra around 110~$\mu$m and of
other broad features such as the one around 120~$\mu$m in NGC 6303
which are absent in the spectrum reduced with our pipeline. The
correction of transients also reduces the high-frequency noise of the
spectrum. In Figure~{\ref{fig:rangecomp}}, below the spectra we show
the ratio of the high-frequency noise in the spectrum produced by the
two pipelines.  In one case the SPG produces a spectrum typically 20\%
noisier than our pipeline. In the other case, depending on the region,
between 10\% and 40\% noisier.  Finally, although the shape and
strength of most of the lines from the two pipelines is very close, we
stress that in the case of bright lines the more aggressive
deglitching used by the SPG pipeline can lead to the removal of the
line peak.  As shown in Figure~{\ref{fig:rangecomp}}, in the case of
Mon R2 the important [CII] line has the peak masked in the archival
products. This reduces the measured flux by approximately 30\%.

\subsection{Mapping}

\begin{figure}
  \vbox{
    \includegraphics[width=\columnwidth]{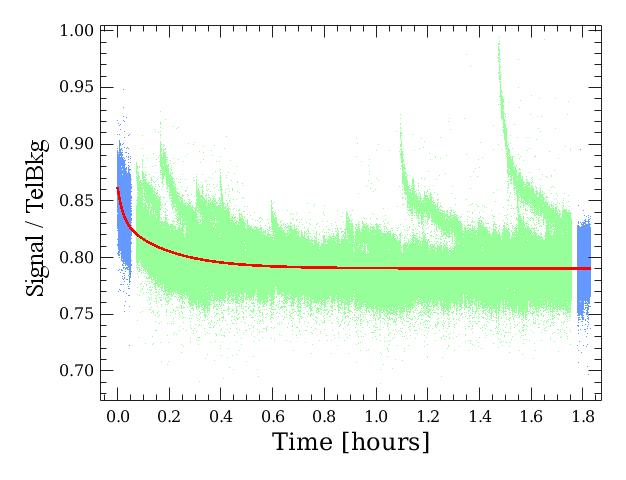}    
  \hbox{
    \includegraphics[width=\columnwidth]{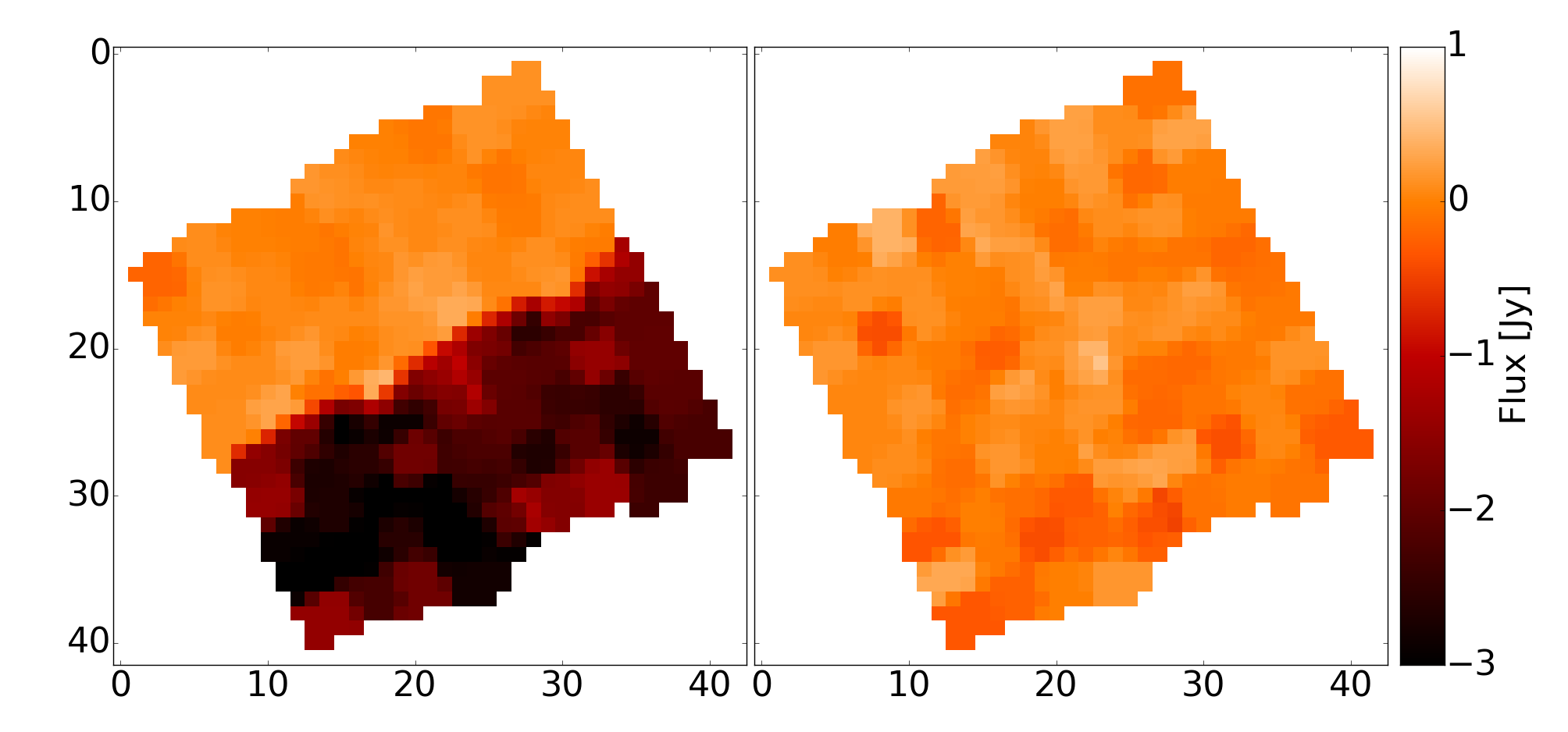}    
  }
  }
  \caption{
    {\it Top panel:} long-term transient for the module 14 of  the observation  1342246963.
    Blue and green points are frames taken off- and on-target.
    {\it Bottom panels:}
    Images at 122~$\mu$m as
    reduced by archive pipeline (left) and the interactive pipeline
    with transient correction (right). The huge transient in this 2$\times$2
    raster observation causes, if not corrected, an uneven background
    with a negative flux in half of the archival image.
  }
  \label{fig:gradient}
\end{figure}

Another effect due to the presence of transients in the data is
clearly visible when considering raster maps. When examining the
observation obsid 1342246963 (see Figure~\ref{fig:gradient}), we notice the
presence of a transient all along the observation.  In particular, the
most affected part is the first off-target observation (blue points in
the figure).  If this transient is not corrected and the off-spectrum
is computed by averaging the two off-target position spectra, the
result is an off-target spectrum with fluxes typically greater than
those of the on-target spectrum. For this reason, after the
subtraction, the flux values are negative.  Moreover, as shown in the
figure, the transient affects the first half of the on-target
observation.  If this is not corrected, the result is a gradient in
the final image. The improvement can be easily seen by eye at the bottom of
Figure~\ref{fig:gradient}, where we compare the image at 122~$\mu$m without (left), and with (right),
the application of the transient correction.

\subsection{Direct comparison with chop-nod}
\label{sec:comparison}

In order to validate the transient correction pipeline release with
version 14 of the HIPE software, a series of observations obtained
with chop-nod and unchopped mode were compared.  The comparison was
limited to the features in the spectra, since the absolute value of
the continuum cannot be obtained with the same accuracy as in the
chop-nod mode.  Only in the case of negligible signal from the source
we can apply a general transient correction to the entire observation
(see e.g. Figure~\ref{fig:ltt}). In all the other cases, strong
continuum signals can affect the solution.  In these cases, we cannot
guarantee the accuracy of the absolute value of the continuum.

\begin{figure}
  \vbox{
    \hbox{
      \includegraphics[width=0.57\columnwidth]{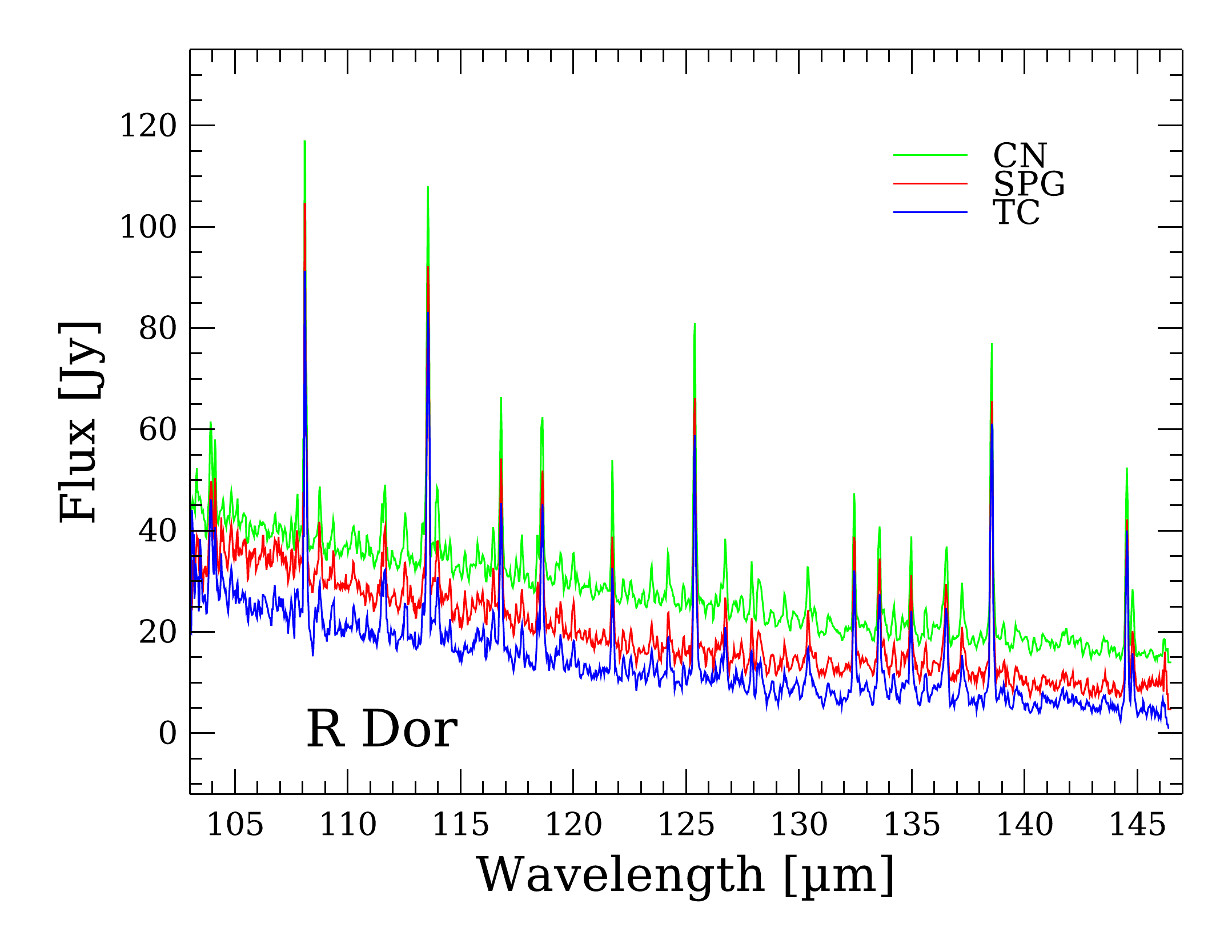}
      \includegraphics[width=0.43\columnwidth]{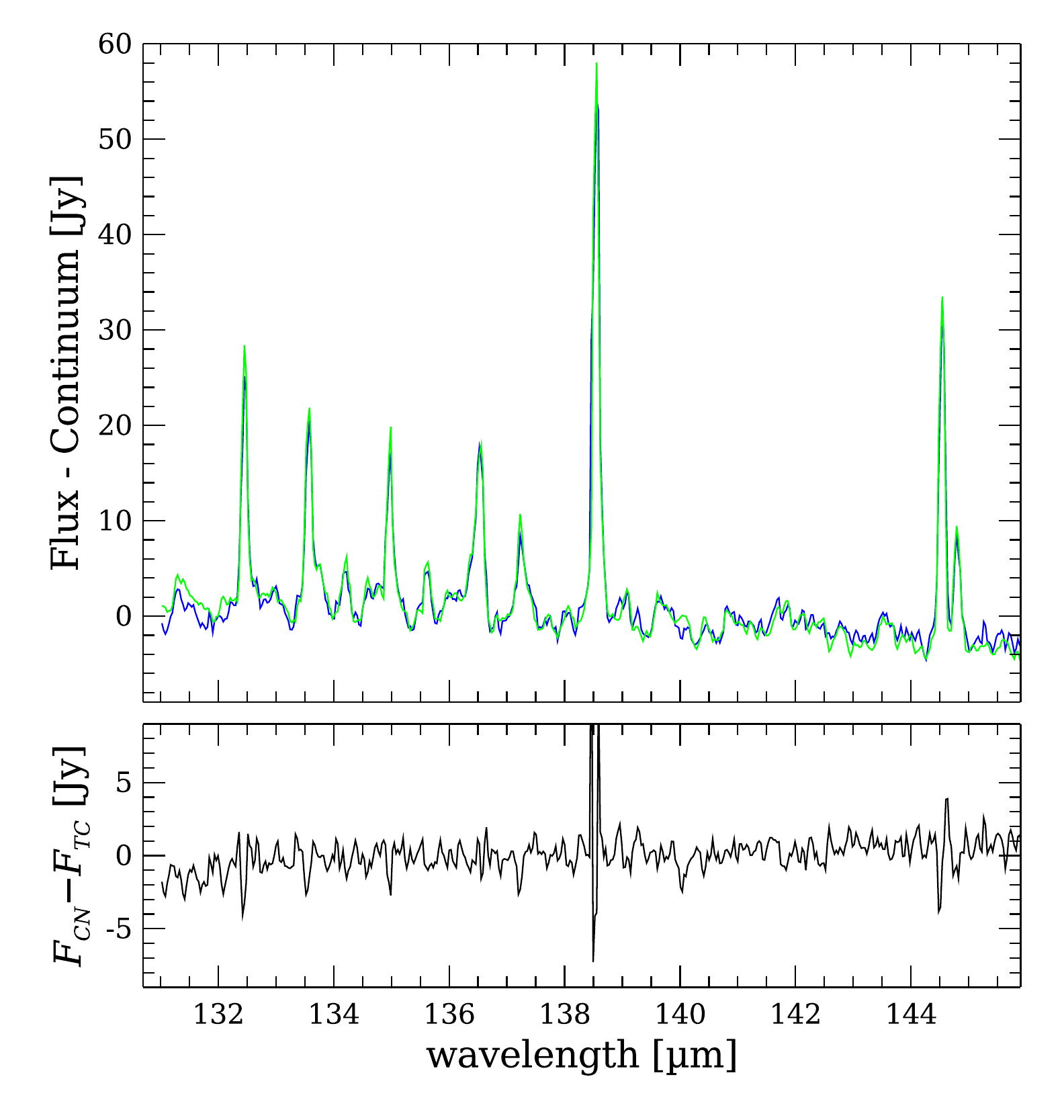}
    }
    \hbox{
      \includegraphics[width=0.57\columnwidth]{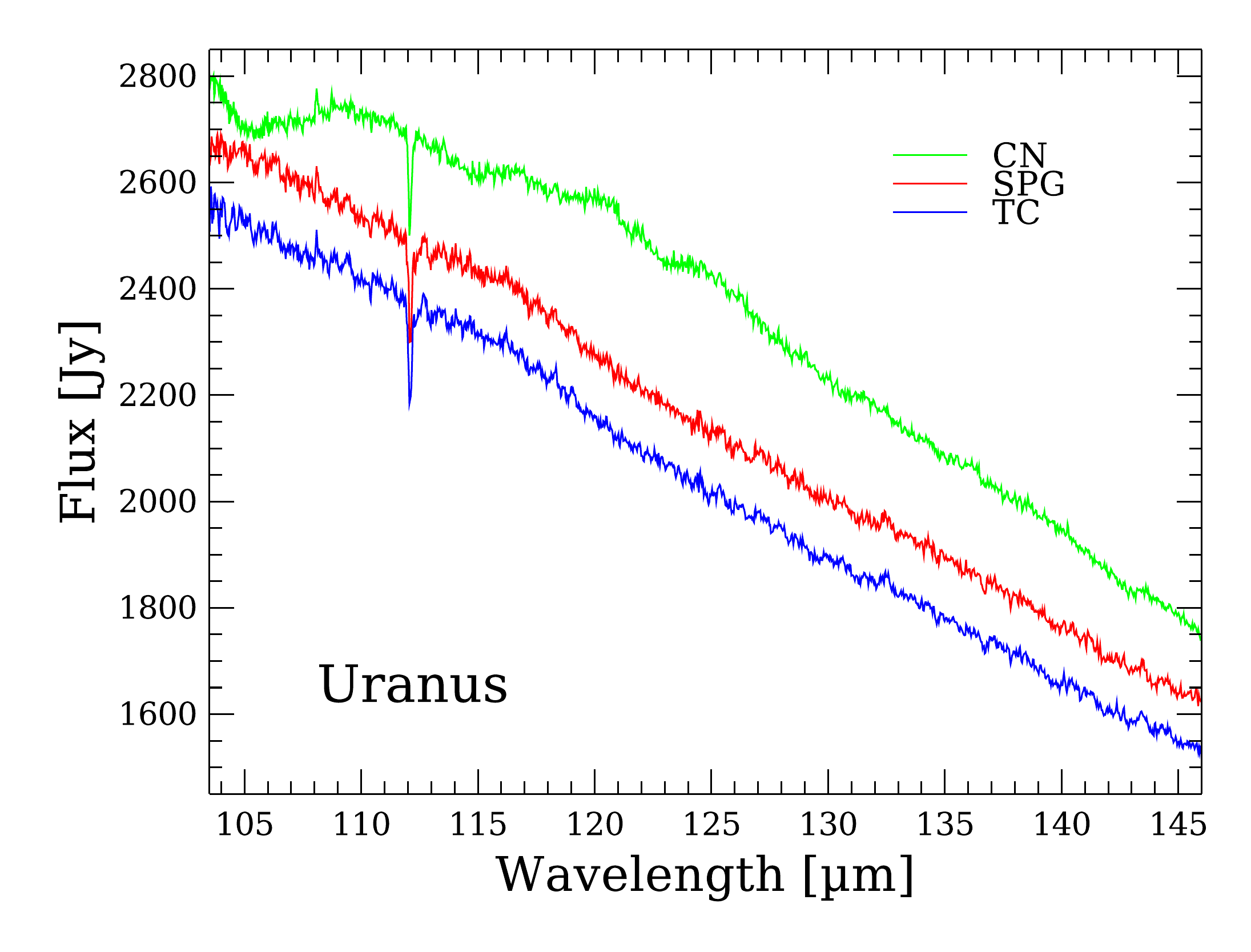}
      \includegraphics[width=0.43\columnwidth]{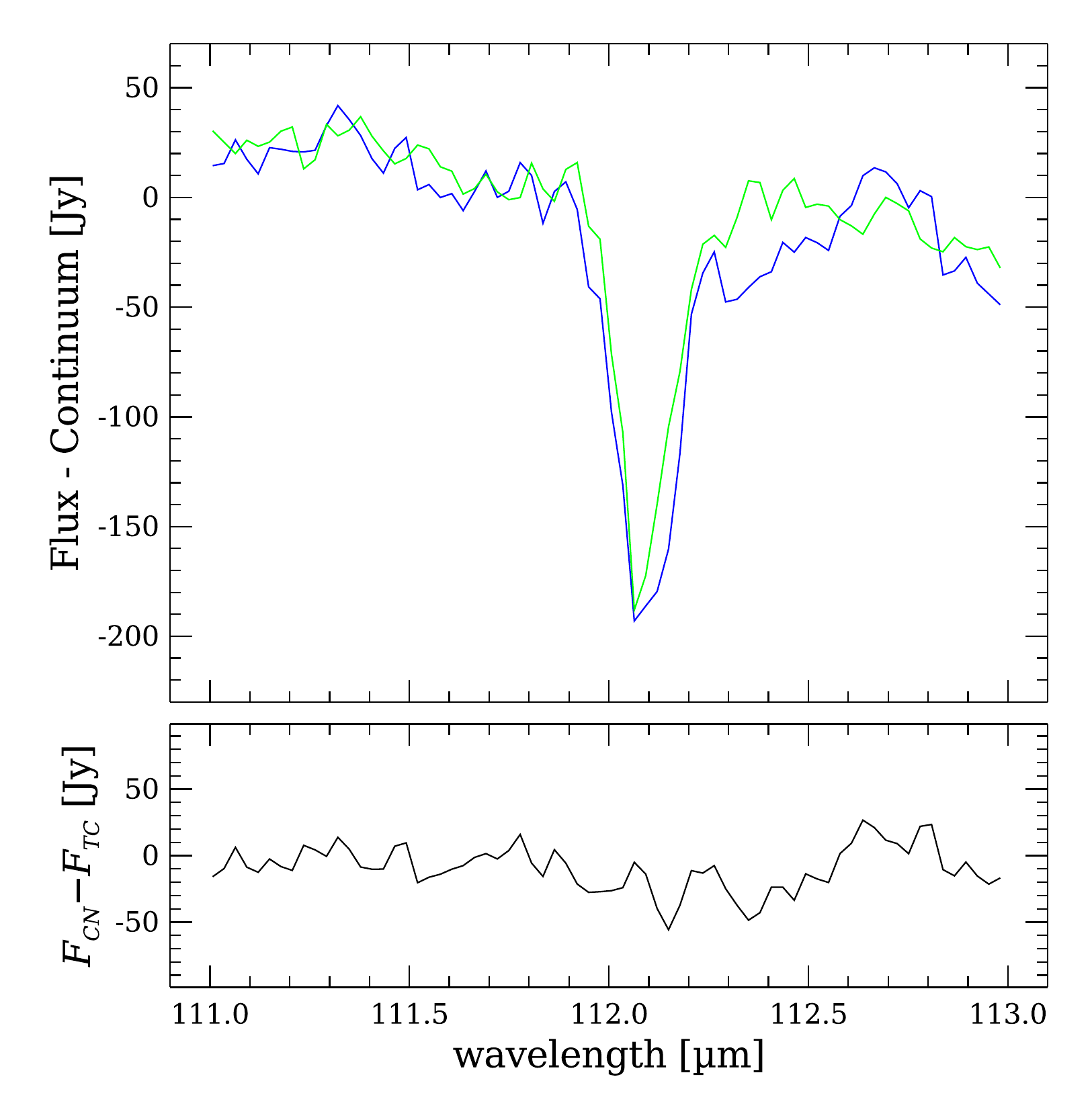}
    }
  }
  \caption{ Comparison between spectra obtained with unchopped (red for SPG, blue for our pipeline) and
    chop-nod (green) modes for R Doradus and Uranus. The right panels show a comparison of emission and absorption lines
    in unchopped and chop-nod modes.
  }
  \label{fig:chopVsUnchop}
\end{figure}

As a first example we considered the observations of the star
R~Doradus.  The top left panel of
Figure~\ref{fig:chopVsUnchop} compares the chop-nod (green, AOR 1342229701) with the
unchopped data (AORs 1342229704/5) from the SPG pipeline (red) and our pipeline (blue).
In the right panel, the reduction with the transient-correction
pipeline is compared to the chop-nod data showing an excellent
agreement.
  We notice that, thanks to the transient correction, the
slope of the spectrum reduced with our pipeline is similar to that of
the chop-nod observations, while the archival spectrum is steeper.

The second example considered is the planet
Uranus\footnote{Incidentally, the planet Uranus was discovered by
  William Herschel himself in 1781, 228 years before the launch of
  {\it Herschel}.}, an extended source on the scale of the PACS
spectrometer spaxels.
We show in the bottom panels of
Figure~\ref{fig:chopVsUnchop} the comparison between the chop-nod (AOR
1342257208) and unchopped observations (AORs 1342257211/2).
Since the
source is extended, we plotted the sum of the $3 \times 3$ central spaxels.
Probably because of pointing problems, the chop-nod data show a turn-around
around 110~$\mu$m which is not present in the unchopped data and it is
not predicted by models \citep[see, for instance,
][]{2014ExA....37..253M}. Also in this case, when directly comparing
the absorption feature of the spectrum, the match is very convincing.

\section{Programs}
In this section we describe the modules written to implement the
various algorithms described in the previous sections, the two
interactive pipelines available in the HIPE distribution for line and
range observations, and the implementation of multi-threading for
speeding up the data reduction which was introduced in HIPE for
transient correction.

\begin{figure*}[t]
  \hbox{
  \includegraphics[width=\columnwidth]{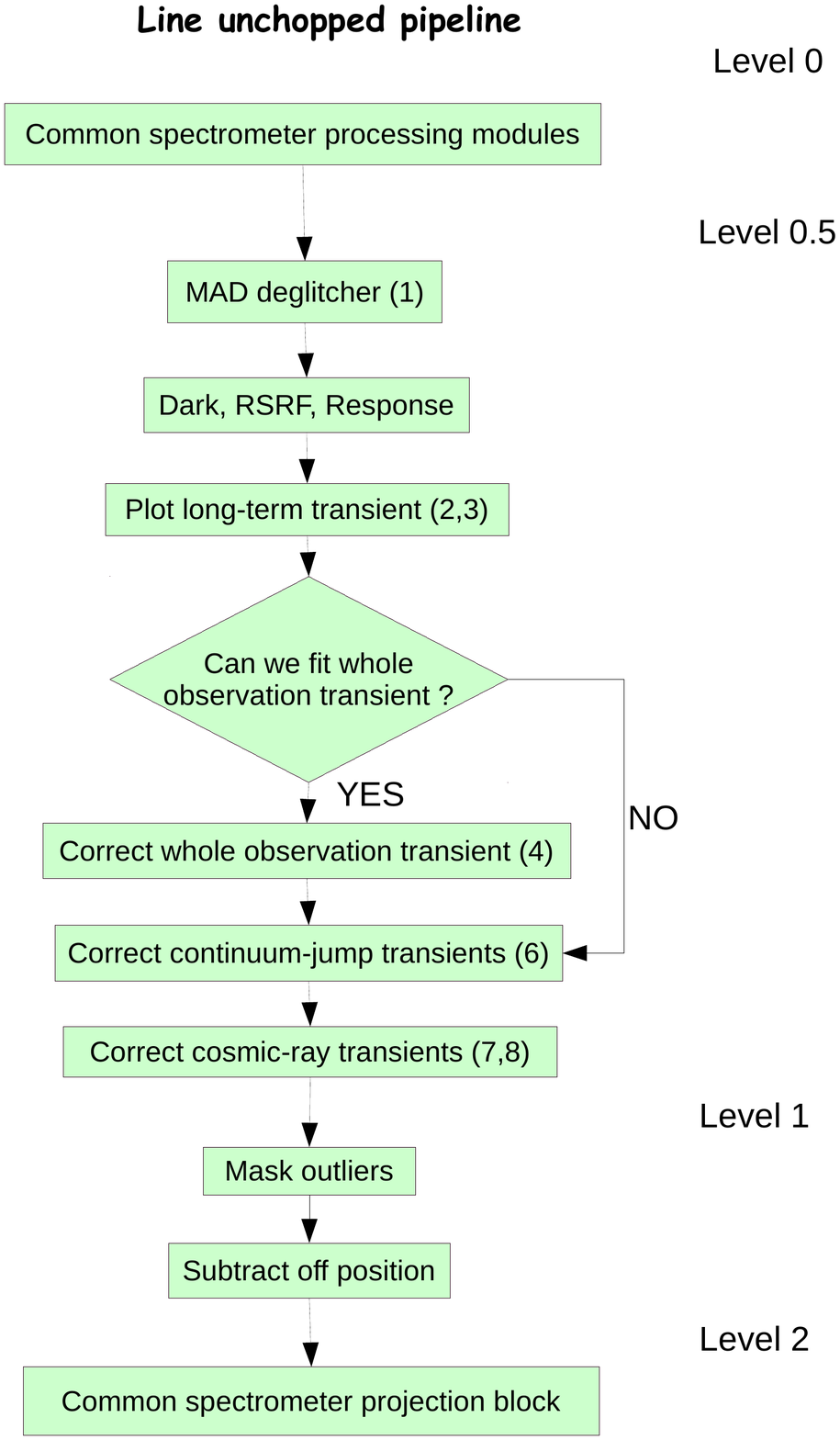}
  \includegraphics[width=\columnwidth]{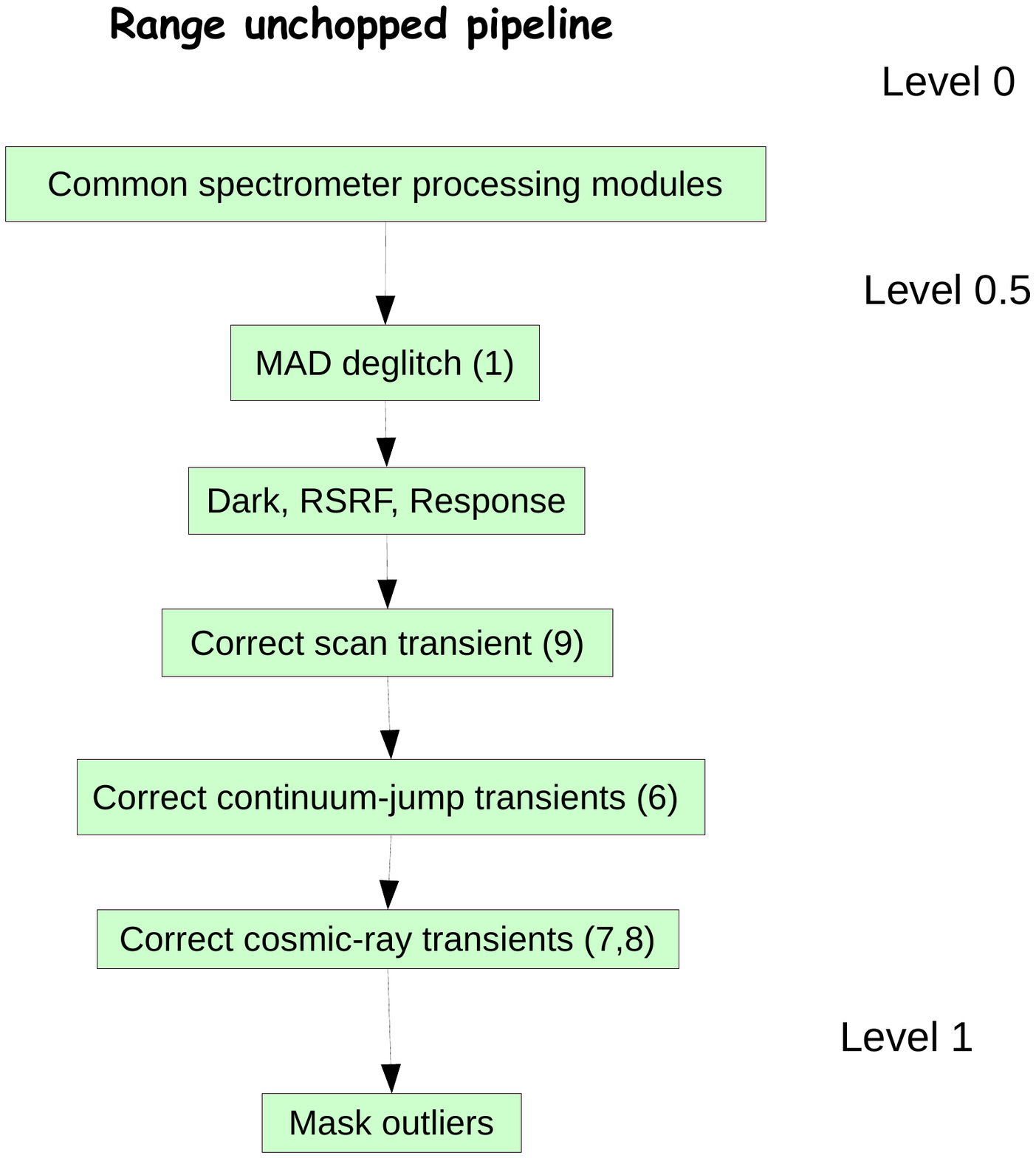}
  }
  \caption{ Flow-charts of the unchopped pipelines for the line and
    range modes (left and right, respectively).  In the range mode the
    pipeline stops at level 1 products. In fact, two observations are
    required to complete the data reduction since the on-target and
    off-target observations are executed in different AORs.
    The numbers in brackets correspond to the tasks used for the step
    as described in section~\ref{sec:modules}.}
  \label{fig:flowcharts}
\end{figure*}

\subsection{HIPE modules}
\label{sec:modules}

Several tasks have been developed to implement the different
algorithms described in the text. Their HIPE names with the numbers
used in Figure~\ref{fig:flowcharts} are:

\begin{enumerate}
\item {\it specFlagGlitchFramesMAD}
\item {\it plotLongTermTransientAll} 
\item {\it specLongTermTransientAll}
\item {\it specApplyLongTermTransient}
\item {\it plotTransient} 
\item {\it specLongTermTransCorr} 
\item {\it specMedianSpectrum} 
\item {\it specTransCorr}
\item {\it specUpDownTransient}
\end{enumerate}

The complete description of these tasks is available in the section 6
of the PACS Data Reduction Guide for spectroscopy.  The task {\it
  specFlagGlitchFramesMAD} replaces the deglitching task {\it
  specFlagGlitchFramesQTest} used in the chop-nod pipeline. The
chop-nod task masks glitches and part of the resulting transient. In
our case, we want to preserve these data to fit our transient
model. We achieve this result by computing the local noise and masking
only outliers beyond a given threshold.  The task {\it
  specLongTermTransientAll} fits the transient along the whole history
of the pixel (see Figure~\ref{fig:ltt}). The fits of transients during
science blocks are performed by the task {\it specLongTermTransCorr}
(see Figure~\ref{fig:lttblock}).  The task {\it specMedianSpectrum}
evaluates the ``Guess'' spectrum (see Figure~\ref{fig:guess}).  An
important bug in this task has been corrected in HIPE 15 (build no. 2845).  Transients
caused by cosmic ray hits are treated with the task {\it specTransCor}
(see Figures~\ref{fig:expected} and~\ref{fig:filter}). This task creates
a new mask (UNCORRECTED) which contains the part of the signal
which cannot be fitted by our model since the interval between
two consecutive discontinuities contains less than 10 points.
Finally, the correction of scan dependent transients described in
section~\ref{sec:scantransients} is implemented in the task {\it
  specUpDownTransient}.

The unchopped transient correction pipelines make use of MINPACK, a
robust package for minimization. MINPACK is more robust in its
handling of inaccurate initial estimates of the parameter values than
other conventional methods. We implemented MINPACK in HIPE as the
``MinpackPro Least Squares Fitting Library''.  This implementation is
based on the original public domain version by
\citet{1980ANL....80...74M}~\footnote{http://www.netlib.org/minpack}.
Our JAVA version is based on a translation to the C language by
S. Moshier~\footnote{https://heasarc.gsfc.nasa.gov/ftools/caldb/help/HDmpfit.html}.
MinpackPro contains enhancements borrowed from C. Markwardt's IDL
fitting routine MPFIT~\footnote{http://purl.com/net/mpfit}
\citep[see also ][]{2009ASPC..411..251M}.

\subsection{HIPE pipelines}

The flow-chart of the pipelines for the reduction of line and range
unchopped data is shown in Figure~\ref{fig:flowcharts}.  In the case
of the line mode, one AOR contains the on- and off-target
observations.  So, the pipeline goes  from the raw data to the final
projected spectra.  We suggest to set the parameter ``interactive'' to
``True'' in order to trigger the interactive step described in the
flow-chart to check if the transient across the entire observation is
correctable. This parameter is set to ``False'' in the script to allow
for automatic checks of the pipeline during software updates.

In the case of the range mode, the main difference is the presence of
the transient correction due to the rapid change of the telescope
background during a wavelength scan.  Also, the correction of the
whole observation transient is not necessary since range-mode AOR
contain only a on- or off-target observation.  For this reason, the
range-mode pipeline stops at level 1 products. To obtain level 2
products, one has to combine the reduction of the on- and off-target
AORs using the ``combine on-off'' script available under the pipeline
unchopped range scan menu.

\subsection{Multi-threading}

Since transient correction tasks are computationally intensive, we
created the multi-threading framework {\it ThreadByRange} in HIPE to
exploit the increasing availability of multiple cores in modern
computers. This multi-threading framework takes advantage of the
organization of the PACS data. In the pipeline, data are stored as
{\it Frames}, i.e. time ordered sets of signal, wavelength, etc.  for
the 25$\times$16 pixels. {\it Frames} are then organized in blocks
called {\it Slices} which correspond to the different parts of an
observation (e.g. the calibration block, off-target observations,
several raster positions, etc.).  The correction tasks can run on each
spaxel independently, so that it is possible to run 25 parallel
processes. On the top of that, each {\it Slice} can be processed
independently.  Therefore, we organize the threads into two pools of
threads for processing. One pool is for threads dedicated to
processing the {\it Slices}, the other to operate on each
spaxel. Resource usage can be controlled by specifying the pool sizes
as task parameters. The default is to create two threads for
processing {\it Slices} and a number of threads for the spaxels equal
to half of the CPUs available.

The tasks implemented in HIPE require a combination of serial and
parallel processes. Taking the signal from or applying the correction
to {\it Frames} must be done serially, since the results of
asynchronous updates to shared objects are undefined. Three methods
are defined in {\it ThreadByRange} to interact with data:
\begin{itemize}
\item {\it preApply()}
  Initialization that must be done serially for the computational
  unit. In the case of pipeline tasks, this involves copying signal,
  wavelengths, masks, etc. to memory that is accessed exclusively by the
  {\it apply()} method.
\item {\it apply()}  {\it ThreadByRange} schedules this method to
  be applied in parallel.  All of the work applied to spaxels
  in parallel is done here, such as minimizations, discontinuity
  detection, application of corrections, etc.
\item {\it postApply()}
  {\it ThreadByRange} calls this method after the completion of each work
  unit.  Transient correction tasks use this method to update {\it Frames}
  with the correction for each spaxel.
\end{itemize}

An absolute requirement for multi-threading is thread safety of the
classes defined in the transient correction tasks. Fortunately, the
level of thread safety was easy to determine due to the robust, clean
design of the numerical package of HIPE.

Our multi-thread framework {\it ThreadByRange} sits on the Java
Concurrency Package from which we use the JAVA {\it
  ExecutorCompletionService} for synchronization, and the {\it
  Executors’ fixedThreadPool} for scheduling.
In the processing of spaxels or {\it Slices}, a {\it Callable} is
created for each computational unit and is submitted to the {\it
  CompletionService}.
This service synchronizes on task completion by using
the {\it BlockingQueue} to retrieve completed tasks as {\it Futures}.
For a complete description of the Java classes 
{\it Callable}, {\it Executor}, {\it BlockingQueue}, {\it Futures}, 
and of this mechanism see
\citet[][section 6.3.5 “Java Completion Service:
Executor meets BlockingQueue”]{2006jcip.book.....G}.

\section{Summary and conclusion}

We presented a description of transient effects on the response of the
PACS spectroscopy detectors.  Taking into account these effects is
paramount in the case of observations using the unchopped modes. In fact,
contrary to the chop-nod mode, there is no way to cancel these effects
by constantly monitoring an off-target signal since on-target and off-target
observations are made at different times.

We showed how it is possible to disentangle and treat separately
transients due to different effects and how to improve dramatically
the signal-to-noise ratio of the final products with respect to those
from the Herschel archive.  In particular, in the line mode the
signal-to-noise ratio of lines can easily triple.  In the case of
range-mode, the correction of transients can lead to changes of slope,
smoother spectra, and better reconstructions of bright lines. The
algorithms described in the paper have been implemented in programs
available in HIPE since version 14, as part of the drop-down menu for
so-called interactive (ipipe) scripts. In particular, two different
pipelines scripts are available for the reduction of line and range
unchopped modes. An important bug affecting the
range unchopped mode was corrected in HIPE version 15 (build 2845).

\begin{acknowledgements}
  The Herschel spacecraft was designed, built, tested, and launched
  under a contract to ESA managed by the Herschel/Planck Project team
  by an industrial consortium under the overall responsibility of the
  prime contractor Thales Alenia Space (Cannes), and including Astrium
  (Friedrichshafen) responsible for the payload module and for system
  testing at spacecraft level, Thales Alenia Space (Turin) responsible
  for the service module, and Astrium (Toulouse) responsible for the
  telescope, with in excess of a hundred subcontractors.

  HCSS and HIPE are a joint developments by the Herschel Science
  Ground Segment Consortium, consisting of ESA, the NASA Herschel
  Science Center, and the HIFI, PACS and SPIRE consortia.

  We are grateful to the entire spectroscopy group of PACS for their
  help and support. In particular, we would like to acknowledge
  P. Royer and B. Vanderbusche for testing the pipeline and pointing
  out significant bugs, as well as A. Poglitsch, R. Vavrek, A.  Contursi, and
  J. de Jong for many useful discussions. We thank K. Exter and the
  anonymous referee for their careful reading of the manuscript and
  very useful suggestions. We would like to thank B. Ali and
  R. Paladini for their constant support at the NASA Herschel Science
  Center. Finally, D.F. is indebted to Prof. I. Perez-Fournon for his
  support at the IAC in a particularly difficult moment of his
  scientific carrier.
\end{acknowledgements}

\bibliographystyle{aa}
\bibliography{manuscript.bib}
 
\end{document}